\documentclass[12pt,aps,pre,amsmath,amssymb]{revtex4}
\usepackage{graphicx}
\usepackage{setspace}

\begin{document}

$\null$
\hfill {July 31, 2017} 
\vskip 0.3in

\begin{center}
{\Large\bf Conformational Heterogeneity and FRET Data}\\

\vskip 0.1in

{\Large\bf Interpretation for Dimensions of Unfolded Proteins}\\

\vskip .5in
{\bf Jianhui S{\footnotesize{\bf{ONG}}}},$^{1,2}$
{\bf Gregory-Neal G{\footnotesize{\bf{OMES}}}},$^{3}$\\
{\bf Tongfei S{\footnotesize{\bf{HI}}}},$^{4}$
{\bf Claudiu C. G{\footnotesize{\bf{RADINARU}}}},$^{3}$
 and
{\bf Hue Sun C{\footnotesize{\bf{HAN}}}}$^{2,*}$

$\null$\\
$^1$ School of Polymer Science and Engineering,
Qingdao University of\\ Science and Technology,
53 Zhengzhou Road, Qingdao 266042, China;\\
$^2$ Departments of Biochemistry and Molecular Genetics,\\
University of Toronto, Toronto, Ontario M5S 1A8, Canada;\\
$^3$ Department of Chemical and Physical Sciences,\\
University of Toronto Mississauga, Mississauga, Ontario L5L 1C6 Canada; and\\
Department of Physics, University of Toronto, Toronto, Ontario M5S 1A7,
Canada;\\
$^4$ State Key Laboratory of Polymer Physics and Chemistry,\\
Changchun Institute of Applied Chemistry, Chinese Academy of Sciences,\\
Changchun 130022, China

\vskip 1.0cm

\end{center}
\noindent
$^*$
Corresponding author.\\
{\phantom{$^*$\ }}Hue Sun Chan.
E-mail: {\tt chan@arrhenius.med.toronto.edu}\\

\vskip 1.8cm

\centerline {\Large To appear in {\it Biophysical Journal}, accepted for
publication}

\vskip 1.3cm

\centerline{\bf
---------------------------------------------------------------------------------------------------------------------}

\noindent
{\tt
Submitted to Biophys J (arXiv: v1): May 16, 2017\\
First decision (Minor Revision): June 14, 2017\\
Revised manuscript submitted (arXiv: v2): July 26, 2017\\
Final acceptance (arXiv: v3 - content same as v2): July 31, 2017
}

\vfill\eject
%\endtitlepage
%-----------------------------------------------------------------------------
%\def\thefootnote{\fnsymbol{footnote}}
%\def\thefootnote{$\dagger$}

\noindent
{\large\bf Abstract}\\

\noindent
A mathematico-physically valid formulation is required
to infer properties of disordered protein conformations
from single-molecule F\"orster resonance energy transfer (smFRET).
Conformational dimensions inferred by conventional approaches that 
presume a homogeneous conformational ensemble can be unphysical. 
When all possible---heterogeneous as well as homogeneous---conformational 
distributions are taken into account without prejudgement, a single
value of average transfer efficiency $\langle E\rangle$ between dyes 
at two chain ends is generally consistent with highly diverse, multiple
values of the average radius of gyration $\langle R_{\rm g}\rangle$. 
Here we utilize unbiased conformational statistics from a 
coarse-grained explicit-chain model to establish a general logical 
framework to quantify this fundamental ambiguity in smFRET inference. 
As an application, we address the long-standing controversy 
regarding the denaturant dependence of $\langle R_{\rm g}\rangle$
of unfolded proteins, focusing on Protein L as an example.
Conventional smFRET inference concluded that $\langle R_{\rm g}\rangle$
of unfolded Protein L is highly sensitive to [GuHCl], but data from 
small-angle X-ray scattering (SAXS) suggested a near-constant 
$\langle R_{\rm g}\rangle$ irrespective of [GuHCl]. Strikingly,
the present analysis indicates that although the reported 
$\langle E\rangle$ values for Protein L at [GuHCl] = 1 M and 7 M are 
very different at 0.75 and 0.45, respectively, the Bayesian $R^2_{\rm g}$ 
distributions consistent with these two $\langle E\rangle$ values 
overlap by as much as $75\%$. Our findings suggest,
in general, that the smFRET-SAXS discrepancy regarding unfolded protein
dimensions likely arise from highly heterogeneous conformational ensembles
at low or zero denaturant, and that additional experimental probes
are needed to ascertain the nature of this heterogeneity.
\\

%%%%%%%%%%%%%%%%%%%%%%%%%%%%%%%%%%%%%%%%%%%%%%%%%%%%%%%%%%%%%%%%%%%%%%%%%%%%%%%
\vfill\eject

%%%%%%%%%%%%%%%%%%%%%%%%%%%%%%%%%%%%%%%%%%%%%%%%%%%
% INTRODUCTION START HERE
%%%%%%%%%%%%%%%%%%%%%%%%%%%%%%%%%%%%%%%%%%%%%%%%%%%

\noindent
{\Large\bf Introduction}\\

\noindent
Single-molecule F\"orster resonance energy transfer (smFRET) 
is an important, increasingly utilized experimental 
technique~\cite{haran2012,schulerCOSB2013,gruebele2014,blanchard2014,rhoades2014,deniz2014,arash2015,rhoades2016,schulerAnnuRev2016}
for studying protein disordered states, especially those of intrinsically
disordered proteins 
(IDPs)~\cite{uversky08,tompa12,julie13,zhirongRev14,cosb15,rohit2015}.
Applications of smFRET to infer conformational 
dimensions of unfolded states of globular 
proteins \cite{haran2006,eaton2007,claudiu2016}
and IDPs \cite{schuler10,baoxu2014,Songetal2015,claudiu2017} 
have provided insights into fundamental protein
biophysics including, for example, folding stability and
cooperativity~\cite{fersht2009,chanetal2011,munoz2012,tobin_forcefields,zhirong2016a},
transition paths~\cite{eaton2013,zhang12}, and compactness 
of IDP conformations \cite{baoxu2014,Songetal2015}
involved in fuzzy complexes \cite{borg07,tanja2010,fuzzy12,Veronika2017}.
Single-molecule conformational dimensions likely bear as well on biologically
functional liquid-liquid IDP phase separation \cite{chongjulie2016} 
because the amino acid sequence-dependent single-chain compactness of
charged IDPs \cite{muthu96,pappu13,kings2015} 
are predicted by theory \cite{linPRL} 
to be closely correlated with these polyampholytic proteins' tendency to 
undergo multiple-chain phase separation \cite{lin2017}. 

Basically, inference from smFRET data on measures of conformational 
dimensions such as radius of gyration $R_{\rm g}$ entails matching 
experimental average energy transfer efficiency $\langle E\rangle_{\rm exp}$ 
with simulated (or analytically calculated) transfer efficiency 
$\langle E\rangle_{\rm sim}$ predicted by 
a chosen polymer model. Using a Gaussian chain model or an augmented
Sanchez mean-field theory, conventional smFRET inference procedures presume 
a homogeneous conformational ensemble that expands or contracts uniformly 
\cite{eaton2007,haranJACS2009,SchulerJCP2013} in response to changes in 
solvent conditions such as denaturant concentration \cite{kiefhaber2006}.
Such an interpretation of smFRET data stipulated a significant collapse of 
unfolded-state conformations, as quantified by a substantial decrease 
in $R_{\rm g}$, upon changing solvent conditions from strongly unfolding to 
folding by lowering denaturant concentration \cite{haran2006,eaton2007}.
This smFRET prediction has led to a long-standing puzzle for 
Protein L~\cite{haran2012,tobinkevinJMB2012,tobinkevinPNAS2015,DanRohit2017} 
because for this two-state folder \cite{Plaxco1999}, an apparently 
more direct measurement of $R_{\rm g}$ by small-angle X-ray scattering (SAXS) 
indicated that the average compactness of its unfolded-state 
conformational ensemble does not vary much with 
denaturant~\cite{haran2012,tobinkevinJMB2012}. Similar behaviors have
also been observed in SAXS experiments on other proteins~\cite{tobin2004}.

Although the smFRET-SAXS puzzle remains to be fully resolved, several 
advances since the discrepancy was first noted~\cite{haran2006}
have contributed to clarifying the pertinent issues. A study 
using explicit-chain models questioned the general 
validity of conventional ``standard'' smFRET interpretation by showcasing
that it incurs substantial errors in inferred $R_{\rm g}$~\cite{dt2009}.
A systematic analysis of subensembles of self-avoiding chains
pinpointed the conventional procedure's basic shortcoming in 
always presuming a homogeneous ensemble, an assumption positing
particular forms of one-to-one mapping between average 
$\langle R_{\rm g}\rangle$ and end-to-end distance $\langle R_{\rm EE}\rangle$
that lead to grossly overestimated $R_{\rm g}$'s for small 
$\langle E\rangle_{\rm exp}$ values~\cite{Songetal2015}. In reality,
however, as should be obvious from polymer theory and explicit-chain 
simulations of polymers, there is no general one-to-one mapping between 
$\langle R_{\rm g}\rangle$ and $\langle R_{\rm EE}\rangle$ if a 
homogeneous ensemble is not assumed, because there are significant scatters 
in the $R_{\rm g}$--$R_{\rm EE}$ relationship (see, e.g., Fig.~2 of 
Ref.~\cite{Songetal2015}).  Therefore, $\langle R_{\rm EE}\rangle$ 
cannot be a proxy 
for $\langle R_{\rm g}\rangle$ in general.
When conformational heterogeneity is recognized, as it is clearly observed in 
a number of smFRET experiments~\cite{kellner2014,claudiu2016}, our
subensemble analysis prescribes a ``most probable'' radius of 
gyration, $R^0_{\rm g}$, for any given 
$\langle E\rangle_{\rm exp}$~\cite{Songetal2015}. The same analysis
shows that $R^0_{\rm g}$ can also correspond to the $\langle R_{\rm g}\rangle$ 
of a distribution of $R_{\rm g}$ consistent with the given
$\langle E\rangle_{\rm exp}$ (Fig.5F of Ref.~\cite{Songetal2015}).
When applied to an N-terminal 
IDP fragment of the Cdk inhibitor Sic1~\cite{borg07,tanja2010,Veronika2017}, 
the subensemble-inferred, denaturant-dependent $R^0_{\rm g}$ is in good 
agreement with SAXS-determined $R_{\rm g}$ and NMR measurement of hydrodynamic 
radius, in contrast to conventional procedures that produced unphysical 
results~\cite{Songetal2015}. 

In line with this conceptual 
framework that emphasizes conformational heterogeneity and
polymer excluded volume, two other recent explicit-chain 
simulation studies also concluded that conventional smFRET 
inference of $R_{\rm g}$ is inadequate~\cite{Reddy2016,zhirong2016b}.
Notably, the coarse-grained model simulation in ref.~\cite{Reddy2016}
predicted an $\approx 3.0$~\AA~contraction of average $R_{\rm g}$ 
for Protein L upon diluting GuHCl from 7.5 M to 1.0 M. 
The authors surmised that $3.0$~\AA~is ``close to the statistical 
uncertainties'' of SAXS-measured $R_{\rm g}$ values, and therefore a 
resolution of the smFRET-SAXS discrepancy for Protein L might be within 
reach~\cite{Reddy2016}. More recently, an extensive experimental-computational 
study of a destabilized mutant of spectrin domain R17 and the IDP ACTR also 
underscored the 
importance of explicit-chain simulations in the interpretation of smFRET 
data. Denaturant-dependent expansion of conformational dimensions was
consistently observed for these proteins from multiple experimental 
methods as well as in all-atom explicit-water molecular dynamics 
simulations~\cite{best2016,schuler2016}. Protein L, however,
was not the subject of this investigation. 

In view of recent results that apparently affirm an appreciable 
denaturant-dependent $R_{\rm g}$ for unfolded proteins---albeit not as
sharp as posited by conventional smFRET interpretation,
is an essentially denaturant-independent unfolded-state 
$\langle R_{\rm g}\rangle$
as envisioned in the usual picture of cooperative protein folding 
tenable? To address this question, we determined
computationally the distribution of $R_{\rm g}$ consistent with any 
given $\langle E\rangle_{\rm exp}$ and the derived probabilities that 
different $\langle E\rangle_{\rm exp}$'s are consistent with the 
same $R_{\rm g}$'s. Taking an agnostic view as to 
the merits of various experimental techniques, we invoked minimal 
theoretical assumption so as to let experimental data speak for themselves.
For simplicity, we do not consider kinetic effects
in smFRET measurements~\cite{steinberg1978,nau2013,hilser2015}.
Accordingly, our coarse-grained model incorporates only the most 
rudimentary geometry of polypeptide chains, without any 
detailed force field such as those applied in recent smFRET-related 
simulations~\cite{dt2009,Reddy2016,best2016}. By this very
construction, our analysis is unaffected by any known or potential 
limitations of current coarse-grained and atomic force 
fields~\cite{cosb15,DavidShaw2,TaoPCCP,sarah15,sarah17,best2017,shea2017}.
As detailed below, we found that simple conformational statistics dictates
a broad distribution of $R_{\rm g}$ for most 
$\langle E\rangle_{\rm exp}$'s. Among such conditional 
(Bayesian~\cite{presse2017}) 
distributions $P(R_{\rm g}|\langle E\rangle_{\rm exp})$'s for different 
$\langle E\rangle_{\rm exp}$ values, large overlaps exist 
even for significantly different $\langle E\rangle_{\rm exp}$'s.
These results suggest that, even if published experimetal data 
are taken at face value, conceivably the smFRET-SAXS discrepancy can be 
resolved provided sufficient denaturant-dependent conformational 
heterogeneity in the unfolded state is encoded by the amino acid sequence 
of the protein.  Our analysis thus establishes a physical 
perimeter within which future experimental and theoretical smFRET 
analyses may proceed.
\\

\noindent
{\Large\bf Methods}
\\

\noindent
The C$_\alpha$ protein model and the sampling algorithm used here are 
the same as that in our previous study~\cite{Songetal2015}. 
The protein is represented by a sequence of $n$ beads connected by 
C$_\alpha$--C$_\alpha$ virtual bonds of length $3.8$~\AA. The potential 
energy $E=\sum_{i=2}^{n-1}\epsilon_\theta(\theta_i-\theta_0)^2+$
$(1/2)\sum_{i=1}^n\sum_{j=1}^n\epsilon_{\rm ex}(R_{\rm hc}/R_{ij})^{12}$,
where $\epsilon_\theta=10.0 k_{\rm B}T$, $\theta_i$ is the virtual bond 
angle at bead $i$, $\theta_0=106.3^\circ$ is the reference that 
corresponds to the most populated virtual bond angle in the Protein
Data Bank~\cite{levitt1976}, $k_{\rm B}$ is the Boltzmann constant, 
$T$ is the absolute temperature, $\epsilon_{\rm ex}=1.0 k_{\rm B}T$ is the 
model protein's self-avoiding excluded-volume 
repulsion strength, and $R_{ij}=|{\bf R}_j-{\bf R}_i|$ is the 
distance between beads $i,j$, wherein ${\bf R}_i$ is the position
vector for bead $i$. The excluded-volume $(R_{\rm hc}/R_{ij})^{12}$ term is 
set to zero for $R_{ij}\ge 10.0$~\AA. 
As in many protein folding simulations~\cite{chanetal2011}.
we use a hard-core repulsion distance $R_{\rm hc}=4.0$~\AA~for most of the 
analysis presented below, while some results for
$R_{\rm hc}=3.14$~\AA~or $5.0$~\AA~\cite{Songetal2015} are also
utilized to assess the robustness of our conclusions.
 
We conducted Monte Carlo sampling by applying the
Metropolis criterion \cite{MC} at $T=300$ K using an algorithm
described previously~\cite{song13} that assigns equal a priori probability 
for pivot and kink jumps~\cite{stockmayer1962,Lal69}. The acceptance 
rate for the attempted chain moves was $\approx 30\%$. 
The first $10^7$ equibrating attempted moves of each simulation were 
excluded from the tabulation of 
statistics. Subsequently, $10^9$ moves were attempted for each
chain length $n$ we studied to sample $10^7$ conformations for
further analysis.  Values of radius of gyration
$R_{\rm g} = \sqrt{n^{-1}\sum_{i=1}^n |{\bf R}_i-{\bf R}_{\rm cm}|^2}$
(where ${\bf R}_{\rm cm}= n^{-1}\sum_{i=1}^n {\bf R}_i$)
and end-to-end distance $R_{\rm EE}=|{\bf R}_n-{\bf R}_1|$ 
were computed for the sampled conformations to determine the distribution
$P(R_{\rm g},R_{\rm EE})$ of populations
centered at various $(R_{\rm g},R_{\rm EE})$ with
only narrow ranges of variations (bins) around the given
$R_{\rm g}$ and $R_{\rm EE}$ values.

We focus here only on cases in which the dyes are attached to the 
two ends of the protein chain. FRET efficiency for a given conformation 
in the model with end-to-end distance $R_{\rm EE}$ is then calculated by
the formula 
\begin{equation}
E(R_{\rm EE}) = \frac {R_0^6}{R_0^6+R^6_{\rm EE}} \; ,
\label{E_eq1}
\end{equation}
where $R_0$ is the F\"orster radius of the dye. Based on the values 
of $R_0=54\pm 3$~\AA~given by Sherman and Haran \cite{haran2006}
and $R_0=54.0$~\AA~provided by Merchant et al. \cite{eaton2007}
for the Alexa 488 and Alexa 594 dyes employed in their Protein L experiments,
we set $R_0=55$~\AA~in most of the computation for Protein L below.
For any given distribution $P(R_{\rm EE})$, the average FRET
efficiency is given by 
$\langle E \rangle = \int d R_{\rm EE}\; E(R_{\rm EE}) P(R_{\rm EE})$.
The subscripts in the above expressions $\langle E\rangle_{\rm exp}$ and 
$\langle E\rangle_{\rm sim}$ are omitted hereafter for notational 
simplicity when the meaning of the average 
$\langle E\rangle$ is clear from the textual context. 
Protein L is a 64-residue $\alpha/\beta$ protein. 
To account for the added effective chain length due to the two dye linkers,
we used $n=75$ chains to model the unfolded-state conformations of
Protein L. This prescription for the linkers is similar to the 
ten \cite{plaxco2005} or eight \cite{eaton2007} extra residues
used before. In addition to the exemplary computation for Protein L,
simulations were also conducted for several other representative
chain lengths ($n=50$, $100$, $125$, and $150$) and F\"orster radii 
($R_0=50$, $60$, and $70$~\AA) for future applications to other 
disordered protein conformational ensembles.
\\

\noindent
{\Large\bf Results}\\

{\bf Physicality of a subensemble approach to smFRET inference.}
To ensure that smFRET inference takes into account only physically
realizable conformations, we recently indroduced
a systematic methodology to infer a most probable radius of 
gyration $R^0_{\rm g}$ from an experimental $\langle E\rangle_{\rm exp}$ 
by considering subensembles of self-avoiding walk (SAW) conformations with 
narrow ranges 
of $R_{\rm g}$ simulated using an explicit-chain model. For any
such range (bin) centered around an $R_{\rm g}$,
the method provides a conditional distribution 
$P(R_{\rm EE}|R_{\rm g})$ for the end-to-end distance $R_{\rm EE}$. 
An average FRET efficiency $\langle E \rangle(R_{\rm g}) 
= \int d R_{\rm EE}\; E(R_{\rm EE}) P(R_{\rm EE}|R_{\rm g})$
is then calculated. The most probable 
$R^0_{\rm g}$ is determined by  matching $\langle E\rangle_{\rm exp}$ with 
$\langle E \rangle(R_{\rm g})$, viz., by solving the equation 
\begin{equation}
\langle E \rangle(R^0_{\rm g}) = \langle E\rangle_{\rm exp} 
\end{equation}
for $R^0_{\rm g}$ to arrive at $R^0_{\rm g}(\langle E\rangle)$
(wherein the ``exp'' is dropped from the average),
which is the inverse function of $\langle E \rangle(R_{\rm g})$.
As documented before~\cite{Songetal2015,claudiu2016} and
outlined above, by explicitly allowing for unfolded-state conformational 
heterogeneity---which is expected physically~\cite{cosb15,rohit2015},
the subensemble SAW method circumvents the limitations of conventional 
smFRET inferences that presuppose a homogeneous conformational 
ensemble~\cite{haran2006,eaton2007,SchulerJCP2013}. 

Based on the same
conceptual framework, here we approach the question of smFRET inference
from a complementary angle. 
Instead of starting from subensembles with
a narrow range of $R_{\rm g}$ to derive $P(R_{\rm EE}|R_{\rm g})$, then 
$\langle E \rangle(R^0_{\rm g})$ and then
$R^0_{\rm g}(\langle E\rangle)$, here we start from subensembles with
a narrow range of $R_{\rm EE}$ (smallest bin size $=0.5$~\AA, see below), and 
hence a narrow variation of $E$ (i.e., via Eq.~(\ref{E_eq1}), the $E$
values in a narrow range may be taken as a single $E$ value), to
derive distribution $P(R_{\rm g}|R_{\rm EE})$ conditioned 
upon $R_{\rm EE}$. While $P(R_{\rm g}|R_{\rm EE})$
is related to $P(R_{\rm EE}|R_{\rm g})$ by Bayes' theorem, 
$P(R_{\rm g}|R_{\rm EE})$ is of interest 
because it quantifies directly the 
possible variation in conformational dimensions 
when only a single $\langle E\rangle_{\rm exp}$ value is known. This
is because for every single FRET efficiency $E$,
the quantity $P(R_{\rm g}|R_{\rm EE})$ is sufficient to
provide the conditional distribution $P(R_{\rm g}|E)$. 
Then, based on these derived $P(R_{\rm g}|E)$ distributions for all individual 
$E$ values, the $P(R_{\rm g}|\langle E\rangle_{\rm exp})$ distribution 
conditioned upon any value of $\langle E\rangle_{\rm exp}$ averaged
from any underlying distribution $P(E)$ of $E$ can be readily obtained. 
\\

{\bf Estimation of conformational dimensions from FRET efficiency is
highly model dependent because of insufficent structural constraint.}
As an exemplary case, we applied this formulation to Protein L. 
Figure~1 shows considerable discrepancies between SAXS-
(squares) and smFRET-deduced (diamonds) 
$\langle R_{\rm g}\rangle$'s, and that different smFRET inference 
approaches lead to very different pictures of how 
$\langle R_{\rm g}\rangle$ of this protein varies with denaturant 
concentration. For a change in [GuHCl] from $\approx 7$ M to $\approx 2$ M,
conventional inference (diamonds)
yielded large $\langle R_{\rm g}\rangle$ decreases of 
$\approx 9$~\AA~(filled diamonds, ref.~\cite{haran2006}) or
$\approx 5$~\AA~(open diamonds, ref.~\cite{eaton2007}). In contrast,
subensemble SAW methods (circles) stipulate a much milder variation with
respect to [GuHCl].  For the same [GuHCl] change, the
most probable $R^0_{\rm g}$ value decreases by 
$\approx 2$~\AA~(open circles) whereas
the change in root-mean-square 
$\sqrt{\langle R^2_{\rm g}\rangle}\equiv
\{\int dR_{\rm g}\; R^2_{\rm g}P(R_{\rm g}|\langle E\rangle_{\rm exp})\}^{1/2}$ 
conditioned upon the published experimental 
$\langle E\rangle_{\rm exp}$ data
is even smaller: it decreases by $\approx 1$~\AA~(filled circles). 
When [GuHCl] is reduced further from 2 M to 0 M, the total decrease over the 
entire [GuHCl] range is $\approx 5.5$~\AA~for $R^0_{\rm g}$
but merely $\approx 2$~\AA~for $\sqrt{\langle R^2_{\rm g}\rangle}$. 
We computed distributions of $R^2_{\rm g}$ and 
$\sqrt{\langle R^2_{\rm g}\rangle}$ here
because these quantities are determined by SAXS~\cite{tobin2004,saxs2015}.
Our results are essentially unchanged if $\langle R_{\rm g}\rangle$ 
is considered instead (see below).

\begin{figure}[t]
\begin{center}
{\includegraphics[width=80mm,angle=0]{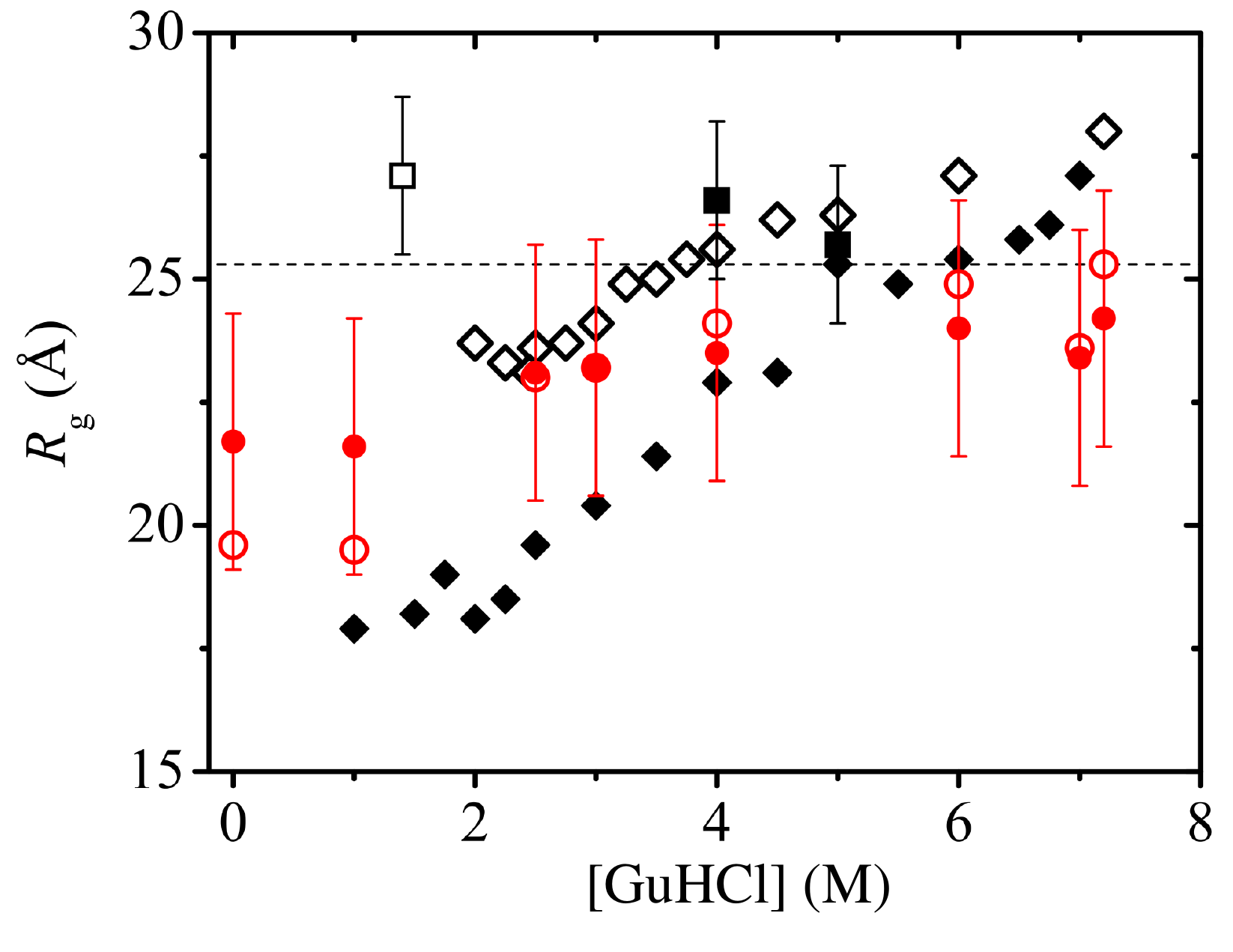}}
\vspace{0.0cm}
\end{center}
\caption{Unfolded-state dimensions of Protein L obtained
from SAXS and various interpretations of smFRET experiments.
Open and filled squares are results from previous time-resolved and
equilibrium SAXS experiments by Plaxco et al. at $2.7\pm 0.5^\circ$C
and $5\pm 1^\circ$C, respectively. The associated error bars
represent one-standard-deviation fitting uncertainties (kinetic data) or
confidence intervals from two to three independent measurements
(thermodynamic data) \cite{Plaxco1999}. Subsequent equilibrium SAXS
measurement at 22$^\circ$C by Yoo et al.~\cite{tobinkevinJMB2012}
produced essentially identical results. Open and filled diamonds
are results from smFRET experiments, respectively, by Merchant et al.
(Eaton group, temperature not provided)~\cite{eaton2007}
and by Sherman and Haran conducted at ``room temperature''~\cite{haran2006}.
These prior experimental data were compared in a similar manner
in ref.~\cite{tobinkevinJMB2012}. Here, the open and filled circles
are from our analysis corresponding, respectively, to the
most-probable $R_{\rm g}^0$ (ref.~\cite{Songetal2015}) and
the root-mean-square
$\sqrt{\langle R^2_{\rm g}\rangle}$ based on the experimental
transfer efficiency $\langle E\rangle=0.74$ for [GuHCl] = 0 given by
Merchant et al., the $\langle E\rangle$ values for Protein L (corrected from
the measured FRET efficiency $\langle E_{\rm m}\rangle$) in
Table~2 of Supporting Information for the same reference~\cite{eaton2007},
and the $\langle E\rangle$ values for [GuHCl] = 1 M and 7 M in
Sherman and Haran~\cite{haran2006}. A F\"orster radius of $R_0=55$ \AA~was
used in our calculations.
The error bars for the open squares span ranges delimited by
$\sqrt{\langle R^2_{\rm g}\rangle \pm \sigma(R^2_{\rm g})}$
where $\sigma(R^2_{\rm g})$ is the standard deviation of the distribution
of $R^2_{\rm g}$ at the given $E$ value. The horizontal dashed line
marks the $R_{\rm g}=25.3$ \AA~ value we obtained from applying the
scaling relation of Kohn et al. \cite{plaxco2004} to $N=74$, where
$n = N+1=75$ is taken to be the equivalent number of amino acid
residues for Protein L plus dye linkers.}
%\label{}
\end{figure}

For every $\langle E\rangle_{\rm exp}$ data point we 
considered for Protein L using subensemble analysis, significant diversity in 
$R^2_{\rm g}$ values that are nonetheless
consistent with the given $\langle E\rangle_{\rm exp}$ 
is observed (Fig.~1, error bars for filled circles). 
In other words, the present method can infer the full Bayesian distribution
of $R_{\rm g}^2$ for a given $\langle E\rangle_{\rm exp}$ and hence
a rigorous error bar can be provided (whereas error bars are 
not provided for $R^0_{\rm g}$ because it represents a narrow range of 
$R_{\rm g}$'s that lead to a distribution of $E$'s which in turn average 
to an $\langle E\rangle$~\cite{Songetal2015}). Figure~1 shows clearly that 
the large variations in inferred $R^2_{\rm g}$ values and the large overlaps of 
the ranges of these variations at different [GuHCl]'s imply that significant 
fractions of the unfolded conformational ensembles of Protein L at different 
[GuHCl]'s can encompass conformations with very similar $R_{\rm g}$'s. 
Notably, the average $R_{\rm g}$ expected of a fully unfolded protein in 
good solvent of the same length as Protein L with dye linkers (horizontal 
dashed line, ref.~\cite{plaxco2004})
is within the $\sqrt{\langle R^2_{\rm g}\rangle}$ error bars
for [GuHCl] as low as 3 M. Even at zero denaturant,
the $R_{\rm g}\approx 24.5$~\AA~value (upper error bar), at
one standard deviation from the mean, $\sqrt{\langle R^2_{\rm g}\rangle}$, 
is only $\approx 1$~\AA~from the average $R_{\rm g}$ expected of a fully 
unfolded conformational ensemble.
\\

{\bf Conformations consistent with a given FRET efficiency generally have
highly diverse radii of gyration.} The diversity in $R_{\rm g}$ values
that are consistent with a given $R_{\rm EE}$ (and therefore a given 
$\langle E\rangle$) is further illustrated in Fig.~2. For our Protein L
model, the square root of the standard deviation in $R_{\rm g}^2$, 
$\sqrt{\sigma(R^2_{\rm g})}$, is substantial for the entire range of
$R_{\rm EE}$: It increases steadily from $\approx 8$~\AA~for
$R_{\rm EE}\approx 0$ to $\approx 12$~\AA~for 
$R_{\rm EE}\approx 120$~\AA~(Fig.~2b). Therefore, although
$\sqrt{\langle R^2_{\rm g}\rangle}$ of the
conformations consistent with a given $R_{\rm EE}$ increases monotonically
from $\approx 18$ to $\approx 37$~\AA~over the $R_{\rm EE}$ range 
in Fig.~2a, knowledge of $R_{\rm EE}$ alone can barely narrow down the
wide range of possible $R_{\rm g}$ values and vice versa (Fig.~2c--f).

\begin{figure}[t]
\begin{center}
{\includegraphics[width=120mm,angle=0]{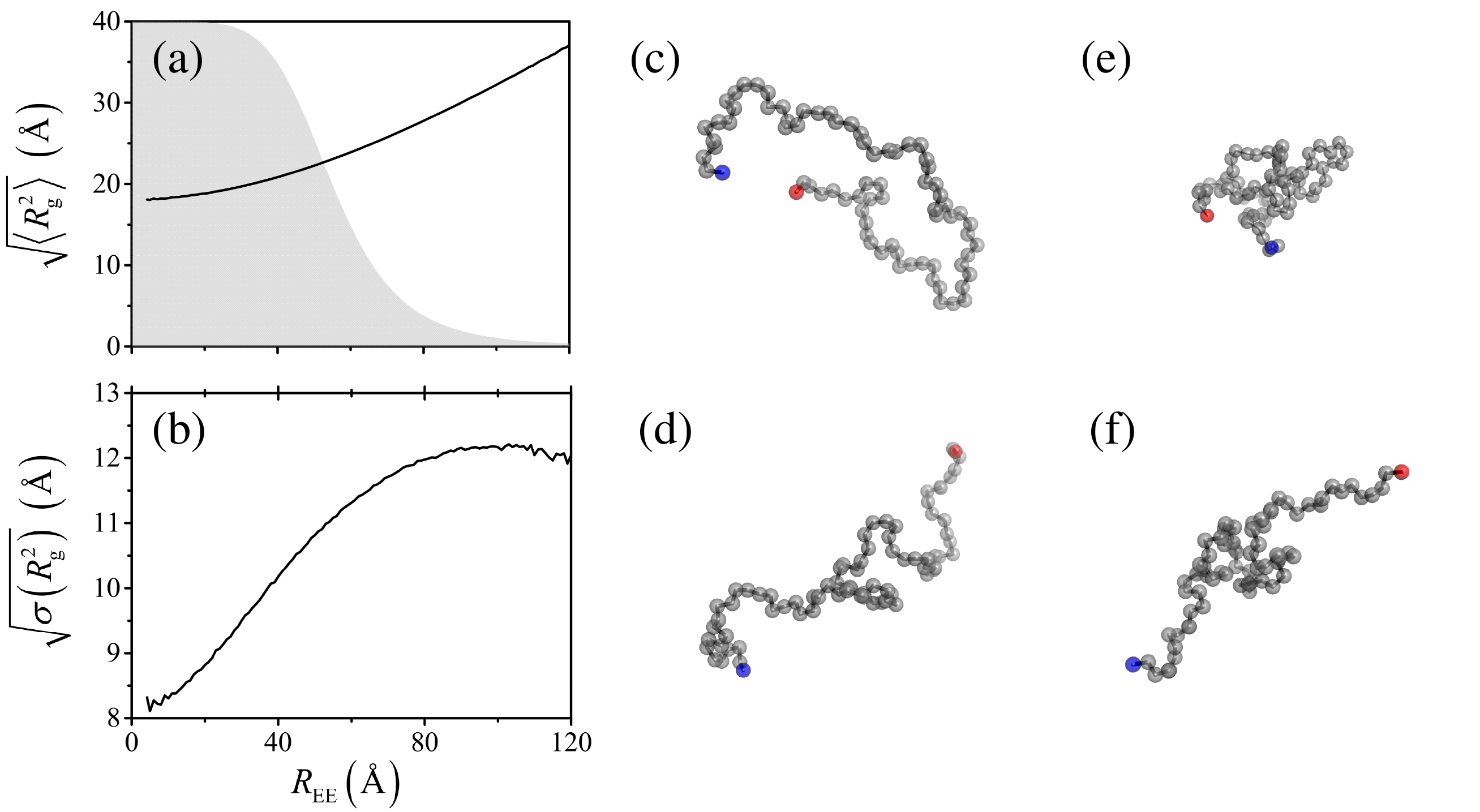}}
\vspace{0.0cm}
\end{center}
\caption{Large variations in dimensions among conformations
with a given end-to-end distance $R_{\rm EE}$.
(a) Root-mean-square $\sqrt{\langle R^2_{\rm g}\rangle}$
and (b) the square root of the standard deviation of $R^2_{\rm g}$
as functions of $R_{\rm EE}$.
The grey profile in (a) shows the theoretical transfer efficiency
Eq.~(\ref{E_eq1}) for $n=75$ and $R_0=55$ \AA~in a vertical scale ranging
from zero to unity.
(c)--(f) Example conformations with the red and blue beads marking the
termini of $n=75$ chains. They serve to illustrate the possible
concomitant occurrences of
(c) small $R_{\rm EE}=19.7$ \AA~ and large $R_{\rm g}=26.3$ \AA;
(d) large $R_{\rm EE}=80.1$ \AA~ and large $R_{\rm g}=26.2$ \AA;
(e) small $R_{\rm EE}=19.7$ \AA~ and small $R_{\rm g}=14.2$ \AA; as well as
(f) large $R_{\rm EE}=80.4$ \AA~ and small $R_{\rm g}=19.8$ \AA. These
examples underscore that there is no general one-to-one mapping
from $\langle R_{\rm EE}\rangle$ to $\langle R_{\rm g}\rangle$.}
\end{figure}

A panoramic view of the logic of smFRET inference on conformational
dimensions is provided by Fig.~3, wherein $P(R_{\rm g},R_{\rm EE})$ 
is converted to $P(R_{\rm g},E)$ by Eq.~(\ref{E_eq1}). Using our model
for unfolded Protein L 
as an example, the landscape in Fig.~3a shows clearly 
that the $R_{\rm g}$--$E$ scatter is wide, with the most populated (red) region 
elongated mainly along the $E$ axis with a small negative incline. 
Consistent with Fig.~1, this population distribution implies
that even large variations in $E$ do not necessitate much
change in the $R_{\rm g}$ distribution. This feature of the 
$R_{\rm g}$--$E$ space is demonstrated more specifically by the 
$\sqrt{\langle R^2_{\rm g}\rangle(E)}$ curve in Fig.~3b 
(red solid curve; the dependence of $\langle R_{\rm g}\rangle$ on 
$E$ is essentially identical, blue solid curve), 
wherein an overwhelming majority of $E$ values are seen to
be consistent with $R_{\rm g}$ values between 20~\AA~and 27~\AA~that are
within one standard deviation of $\sqrt{\langle R^2_{\rm g}\rangle(E)}$
(red dashed curves). In contrast, conventional smFRET inference 
procedures---which are demonstrably unphysical in some 
situations~\cite{Songetal2015}---posit a much more sensitive dependence 
of inferred $\langle R_{\rm g}\rangle$ on $\langle E\rangle$ (Fig.~S1).
It is noteworthy that, for most $E$ values, 
the variation of $\sqrt{\langle R^2_{\rm g}\rangle(E)}$ is 
milder than that of $R^0_{\rm g}(\langle E\rangle)$; i.e.,
$ \vert d\sqrt{\langle R_g^2\rangle}/dE\vert 
< \vert dR^0_g/d\langle E\rangle\vert$.
In fact, this trend is already evident in Fig.~1 from 
the milder [GuHCl] dependence of $\sqrt{\langle R_g^2\rangle}$ 
(filled circles) than that of $R^0_{\rm g}$ (open circles). 
\\

\begin{figure}[t]
\begin{center}
{\includegraphics[width=70mm,angle=0]{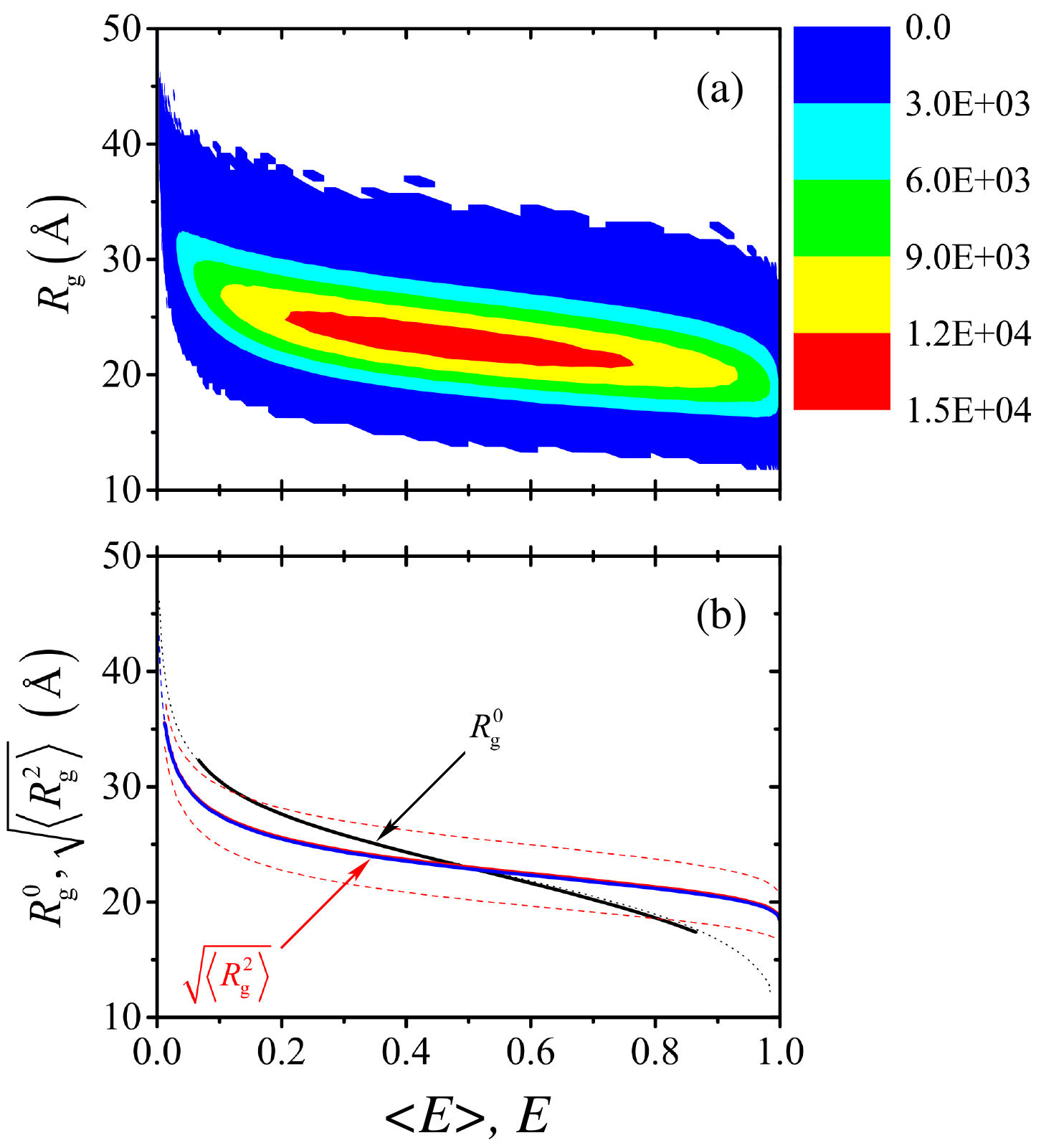}}
\vspace{0.0cm}
\end{center}
\caption{Perimeters of inference on conformational dimensions
from F\"orster transfer efficiency.
(a) Distribution $P(R_{\rm g},E)$ of conformational population as a
function of $R_{\rm g}$
and $E$ for $n=75$ and $R_0=55$ \AA. The distribution was computed using
$R_{\rm EE}\times R_{\rm g}$ bins of $1.0$\AA$\times 0.5$\AA.
White area indicate bins with no sampled population.
(b) Most-probable radius of gyration $R^0_{\rm g}(\langle E\rangle)$ from
our previous subensemble SAW analysis~\cite{Songetal2015} (black solid curve)
compared against root-mean-square radius of gyration
 $\sqrt{\langle R^2_{\rm g}\rangle(E)}$
(red solid curve) computed by considering 30 subensembles with
narrow ranges of $R_{\rm EE}$. The latter overlaps almost completely with
$\langle R_{\rm g}\rangle (E)$ computed using the same set of subensembles
(blue solid curve). Another set of $R^0_{\rm g}(\langle E\rangle)$ values
(black dotted curve) and another set of $\langle R_{\rm g}\rangle (E)$
values (blue dashed curve)
were obtained from the distribution in (a), respectively,
by averaging over $E$ at given $R_{\rm g}$ values and
by averaging over $R_{\rm g}$ at given $E$ values.
Variation of radius of gyration is illustrated by the red dashed curves
for $\sqrt{\langle R^2_{\rm g}\rangle \pm \sigma(R^2_{\rm g})}$
as functions of $E$.
The essential coincidence between the black solid and dotted curves
and between the blue solid and dashed curves indicate that the present
results are robust with respect to the choices of bin size we have made.
Note that the black solid curve for $R^0_{\rm g}(\langle E\rangle)$
does not cover $\langle E\rangle$ values close to zero or close to unity
because larger $R_{\rm g}$ bin sizes ($\sim 1.1$--$3.6$ \AA) than the
current $R_{\rm g}$ bin size of 0.5 \AA~were used
(Table~S5 of ref.~\cite{Songetal2015}), thus precluding extreme values
of $\langle E\rangle$ to be considered in that previous $n=75$ subensemble
SAW analysis~\cite{Songetal2015}. This limitation is now rectified
for $n = 75$ (black dotted curve).}
\end{figure}

{\bf Conformations sharing similar radii of gyration can have very
different FRET efficiencies.} In light of the large diversity
in $R_{\rm g}$ values conditioned upon a given $E$ and
the very mild variation of $\sqrt{\langle R_g^2\rangle}$ and
$\sigma(R^2_{\rm g})$ with $E$ (Fig.~3), one expects that 
conformations consistent with even very different $E$ values 
share highly overlapping $R_{\rm g}$ values. We now characterize this
overlap quantitatively by first considering two sharply defined
representative $R_{\rm EE}$ values in Fig.~4a (vertical bars depicting
$\delta$-function-like distributions) that correspond, by
virtue of Eq.~(\ref{E_eq1}), to two sharply defined $E$ values 
$\approx 0.45$ and $0.75$ (Fig.~4b). These $E$ values are
representative because 
they coincide with the experimental $\langle E\rangle_{\rm exp}$
for Protein L at [GuHCl] = 7 M and 1 M, respectively~\cite{haran2006}.
The conditional distributions
$P(R^2_{\rm g}|E)$ for $E=0.45$ and $E=0.75$ 
overlap significantly, with the overlapping area 
$\approx 0.75$ (Fig.~4c). By definition, this area is the overlapping 
coefficient, OVL, used in statistical analysis for measuring similarity between 
distribution~\cite{ovl1989}. OVL between two distributions
is generally given by
\begin{equation}
{\rm OVL}_{1,2} = \int dx \; \min [P_1(x),P_2(x)] \; ,
\label{ovl_eq}
\end{equation}
where $P_1(x)$ and $P_2(x)$ are two normalized distributions of
variable $x$. The $P_1$, $P_2$ distributions are
$P(R^2_{\rm g}|E=0.45)$ and $P(R^2_{\rm g}|E=0.75)$ in Fig.~4c.

\begin{figure}[t]
\begin{center}
{\includegraphics[width=95mm,angle=0]{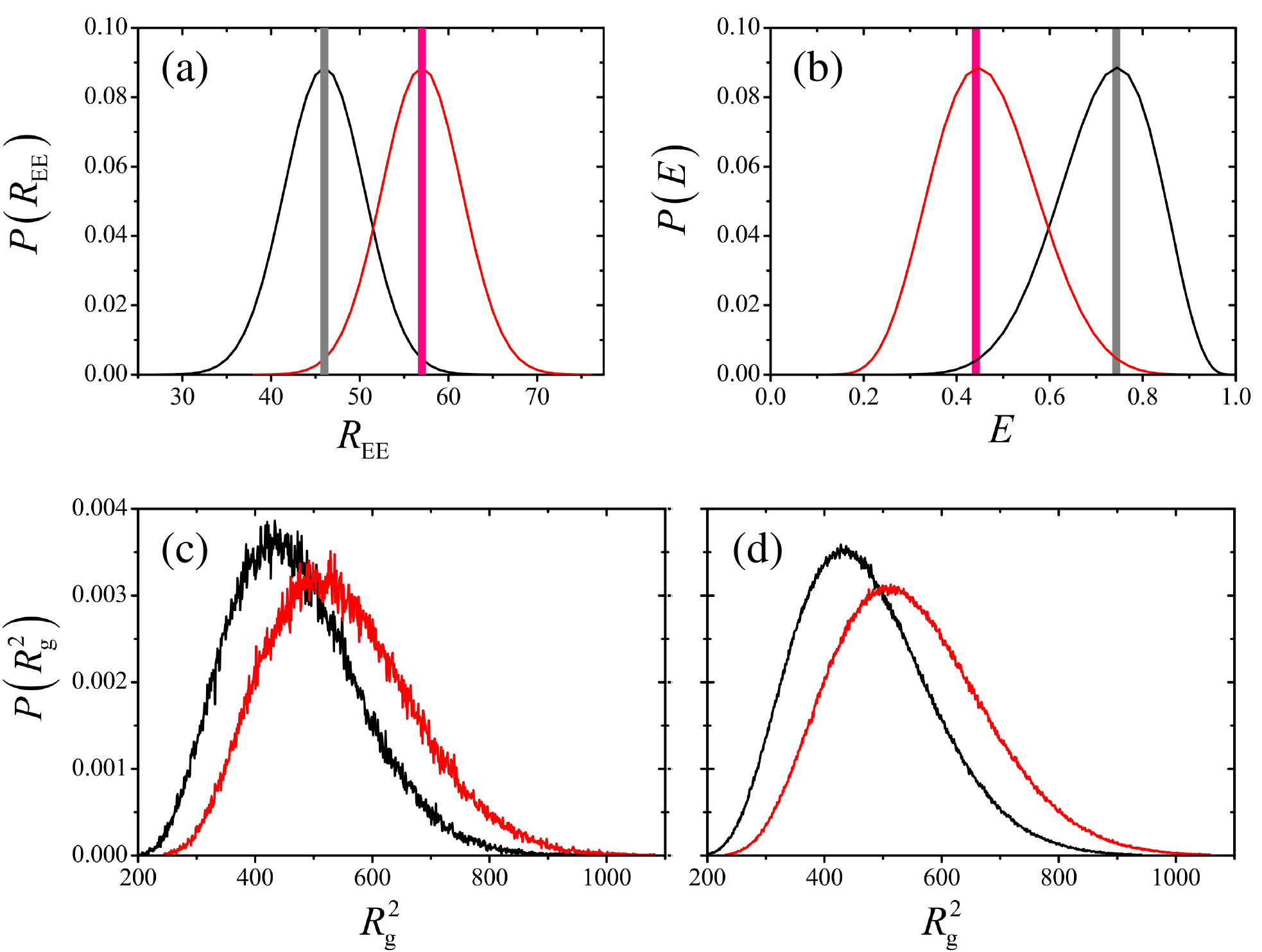}}
\vspace{0.0cm}
\end{center}
\caption{Substantially overlapping distributions of
conformational dimensions can be consistent with very
different F\"orster transfer efficiencies.
(a) Hypothetical distributions $P(R_{\rm EE})$ of end-to-end
distance $R_{\rm EE}$.  Two hypothetical sharp distributions
at two $R_{\rm EE}$ values (vertical bars) and two hypothetical broad
Gaussian distributions (bell curves) centered at these
two $R_{\rm EE}$ values, with standard deviation of the Gaussian
distributions chosen to be $20.3$ \AA.
(b) The corresponding distribution $P(E)$ of F\"orster transfer
efficiency $E$. The left and right sharp distributions of $P(R_{\rm EE})$
in (a) lead, respectively, to $E\approx 0.745$ (right)
and $E\approx 0.447$ (left) in (b). The corresponding $P(E)$ for
the hypothetical
Gaussian distributions in (a) entail broad distributions in $E$ in (b) with
mean values at $\langle E\rangle = 0.735$ (right)
and $\langle E\rangle = 0.453$ (left) respectively.
(c) The left and right curves are the conditional distributions
$P(R^2_{\rm g}|E)$, respectively,
for the sharply defined $E\approx 0.745$ and $E\approx 0.447$ in (b).
(d) Similar to (c) except the distributions of $R^2_{\rm g}$ are
now for the two broad $P(E)$ distributions
in (b). We denote these distributions
as $P(R^2_{\rm g}|\langle E\rangle)$.
The $R^2_{\rm g}$ bin size in (c) and (d) is $1.0$ \AA$^2$.
The overlap area (OVL) of the two normalized distribution curves
in (c) and (d) are, respectively, 0.747 and 0.754.
The percentages of population with $R^2_{\rm g}\ge 625$ \AA~in the
distributions in (c) and (d) are, respectively, 9.2\% and 10.1\% for
$E\approx 0.745$ and $\langle E\rangle=0.735$,
and 25.2\% and 26.3\% for $E\approx 0.447$ and $\langle E\rangle=0.453$.}
\end{figure}

Because experimentally determined $E$ values are often averages, not 
sharply defined~\cite{haran2006,eaton2007}, it is necessary to address 
more realistic distributions of $E$ on smFRET inference.
We do so here by considering hypothetical broad Gaussian distributions 
for $R_{\rm EE}$ centered around the two sharply defined $R_{\rm EE}$ 
values (Fig.~4a, curves, standard deviation $\sigma(R_{\rm EE})=20.3$~\AA), 
resulting in broad distributions in $E$ averaging to
$\langle E\rangle=0.45$ and $0.74$ (Fig.~4b, curves), which are essentially 
equal to the sharply defined $E$ values of $0.45$ and $0.75$.
Modifying the two sharply defined $E$ values to two broad distributions of
$E$ has very little impact on either the individual 
$R^2_{\rm g}$ distributions [$P(R^2_{\rm g}|\langle E\rangle)$]
or the overlap of the two $P(R^2_{\rm g}|\langle E\rangle)$ 
distributions (Fig.~4d). The overlapping coefficient remains $\approx 0.75$. 

Although the distributions 
in Fig.~4c and 4d are very similar, there is a basic
difference between two sharply defined $E$ values and
two broad distributions of $E$ in regard to the conformations 
in the $R^2_{\rm g}$ distributions. When the $E$ values are sharply defined,
there is no overlap in the actual conformations in the 
two $P(R^2_{\rm g}|E)$ distributions 
because the conformational ensembles consistent with 
two sharply defined $R_{\rm EE}$ values are disjoint.
However, when the two sets of $E$ values are broadly distributed
with overlapping $R_{\rm EE}$ and $E$ values (Fig.~4a, b; curves), some 
of the conformations from the two different $R^2_{\rm g}$ distributions
that contribute to the overlapping region in Fig.~4d can be identical.
\\

{\bf The distribution of radius of gyration consistent with a given single
FRET efficiency is very similar to that consistent with a
symmetric distribution of FRET efficiencies centered around it.}
This insensitivity of the distribution of $R^2_{\rm g}$ (and 
therefore also of $R_{\rm g}$) conditioned upon given $E$ values 
to variations in the width of Gaussian-like distribution
of $E$ is not difficult to fathom. Given the
mild variation of $\sqrt{\langle R_{\rm g}^2\rangle}$ and 
$\sigma(R^2_{\rm g})$ with respect to $E$ (Fig.~3b)
and the tendency for effects from $E$ values 
on opposite sides of the average of a symmetric distribution to 
cancel each other, averaging over a range of $E$ values centered around 
a given $E$ ($=\langle E\rangle$) is not expected to result in an overall 
average $R_{\rm g}^2$ and overall distribution width that are substantially 
different from those for a sharply defined $E=\langle E\rangle$.
For the sake of testing the robustness of this insensitivity,
here we have used a large standard deviation, $\sigma(R_{\rm EE})$, 
for the hypothetical Gaussian distributions in Fig.~4a. This 
$\sigma(R_{\rm EE})$ is equal to the standard deviation of the 
$R_{\rm EE}$ distribution for the full conformational ensemble (with the mean,
$\langle R_{\rm EE}\rangle = 59.1$~\AA). Beside the 
$R_{\rm EE}$ and $E$ distributions in Fig.~4, we performed
additional calculations using Gaussian distributions of $R_{\rm EE}$ 
centered at different averages, 
with different standard deviations that equal $0.1\times$, 
$0.25\times$, $0.5\times$, and 
$0.75\times\sigma(R_{\rm EE})$. These constructs
beget distributions of $E$ with different $\langle E\rangle$ values.
In all cases we considered, the resulting $R^2_{\rm g}$ distribution
for the given $\langle E\rangle$
is essentially the same across the different standard deviations as well
as for the case with a sharply defined $E=\langle E\rangle$.
This finding suggests that the $\sqrt{\langle R_g^2\rangle(E)}$--$E$ dependence
in Fig.~3b is not strictly limited to
sharply defined $E$ values. An essentially identical relationship
should also be
is applicable to the $\sqrt{\langle R_g^2\rangle(\langle E\rangle)}$
and associated $\sigma(R^2_{\rm g})$ conditioned upon 
reasonably symmetric distributions of $E$ with mean value $\langle E\rangle$.
In other words, $\sqrt{\langle R_g^2\rangle(E)}$ in Fig.~3, which
was originally constructed for sharply defined $E$ values, is also expected
to be a good approximation of 
$\sqrt{\langle R_g^2\rangle(\langle E\rangle)}$ for
essentially symmetric distributions of $E$. More generally, the
$\sqrt{\langle R_g^2\rangle(\langle E\rangle)}$ for any distribution
$P(E)$ of $E$, symmetric or otherwise, can be 
calculated readily as $[\int dE\; P(E)\langle R_g^2\rangle(E)]^{1/2}$
by using the $\langle R_g^2\rangle(E)$ values from Fig.~3.
\\

{\bf Inference of conformational dimensions solely from FRET efficiency 
can entail significant ambiguity.} To ascertain more generally the degree to
which the $R_{\rm g}$ values consistent with different FRET efficiencies 
overlap, we extended the comparison in Fig.~4c for two $E$ values by 
computing the corresponding overlapping coefficients (Eq.~(\ref{ovl_eq})) 
for all possible pairs of FRET efficiencies, $E_1$ and $E_2$:
\begin{equation}
{\rm OVL}(R^2_{\rm g})_{E_1,E_2}=\int dR^2_{\rm g} \; 
\min [P(R^2_{\rm g}|E_1),P(R^2_{\rm g}|E_2)] \; .
\end{equation}
The heat map in Fig.~5 indicates substantial overlaps for a majority
of $(E_1, E_2)$. Among all possible $(E_1, E_2)$ combinations,
more than 30\% have OVL $\ge 0.8$, and close to 60\% have
OVL $\ge 0.6$ (Fig.~S2a), meaning that their $P(R^2_{\rm g}|E)$'s are
quite similar. Notably, OVL increases significantly 
as $E_1,E_2$ increase above $\approx 0.4$. We also computed averages 
of $R^2_{\rm g}$ over the overlapping regime of the pairs of distributions. 
These averages represent conformational dimensions that are consistent 
with both $E_1$ and $E_2$.  In a majority of 
the situations, the root-mean-square $R^2_{\rm g}$ for the overlapping 
regime stays within a relative narrow range of $\approx 22$--$25$~\AA~for 
our model of unfolded Protein L , even for $E_1$ and $E_2$ that 
are quite far apart (Fig.~S2b). Therefore, taken together with Figs.~1--4, 
the overview in Fig.~5 indicates that when an explicit-chain physical model is 
used to interpret/rationalize smFRET data~\cite{Songetal2015,claudiu2016}, 
as is the case here, the a priori expectation is that 
even substantial changes in $\langle E\rangle_{\rm exp}$
do not necessarily imply large changes in average $R_{\rm g}$. In this light,
previous smFRET-based stipulations of large denaturant-dependent changes 
in the $\langle R^2_{\rm g}\rangle$ of Protein L \cite{haran2006,eaton2007} 
is demonstrably inconclusive in the absence of additional relevant 
experimental information, because they were based on conventional
inference approaches that are not entirely physical~\cite{Songetal2015}.
Moreover, as is evident from the examples in Fig.~6, the trend of a mild 
$R_{\rm g}$--$E$ variation that we saw previously~\cite{Songetal2015}
and in Figs.~1--5 here, which is derived directly from explicit-chain
polymer models, is expected to hold generally for other FRET systems of 
disordered proteins with different chain lengths and F\"orster radii as well. 
\\

\begin{figure}[t]
\begin{center}
{\includegraphics[width=120mm,angle=0]{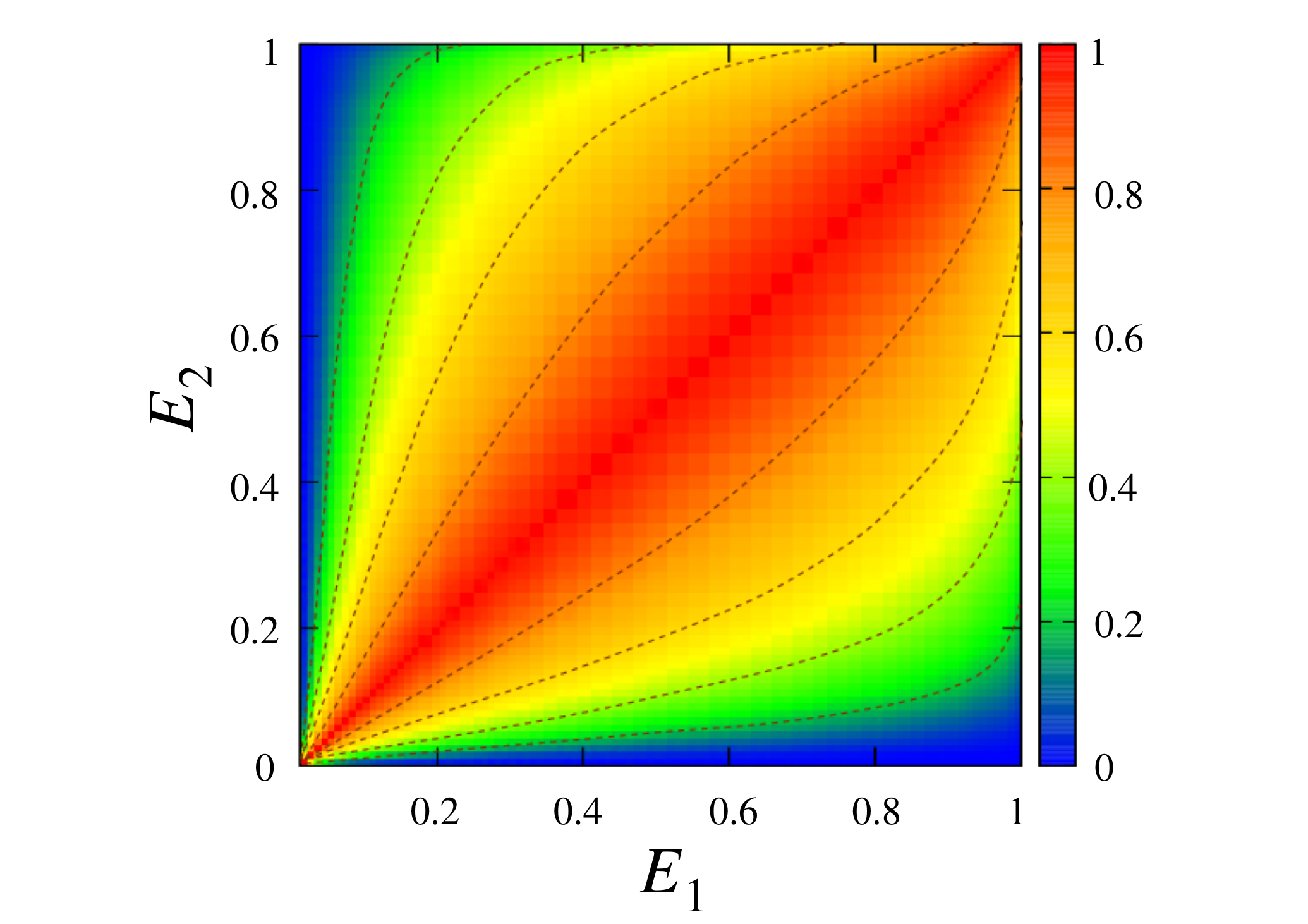}}
\vspace{0.0cm}
\end{center}
\caption{Ambiguities in FRET inference of conformational dimensions.
The heat map provides for $n=75$ and $R_0=55$ \AA~the overlapping coefficient
${\rm OVL}(R^2_{\rm g})_{E_1,E_2}$ of pairs of $R^2_{\rm g}$ distributions
conditioned
upon FRET efficiencies $E_1$ and $E_2$. Contours on the heat map are for
${\rm OVL}(R^2_{\rm g})_{E_1,E_2}$
= 0.8, 0.6, 0.4, and 0.2, as indicated by the color scale on the right.}
\end{figure}

\noindent
{\Large\bf Discussion}\\

{\bf Subensemble-derived conditional distributions of $R_{\rm g}$ 
are basic to smFRET inference.}
To recapitulate, here we have further developed the subensemble SAW
approach to smFRET inference of conformational dimensions~\cite{Songetal2015},
which is based on the obvious principle that only physically realizable
conformational ensembles should be invoked to interpret smFRET data.
We focused previously on the most probable radius of gyration 
$R^0_{\rm g}(\langle E\rangle)$, which is derived from distributions 
of $E$ conditioned upon a narrow range of $R_{\rm g}$. Here we have
considered the complementary quantity, $\sqrt{\langle R^2_{\rm g}\rangle(E)}$,
which is the root-mean-square value of $R_{\rm g}$ conditioned upon a given 
$E$. These quantities are not identical, but their variations with
$\langle E\rangle$ or $E$ are similar (Figs.~3 and 6). Relative
to conventional approaches to smFRET inference, both 
$R^0_{\rm g}(\langle E\rangle)$ and $\sqrt{\langle R^2_{\rm g}\rangle(E)}$
exhibit a milder dependence on smFRET efficiency, covering a range of
$R_{\rm g}$ values consistent with polymer physics~\cite{Songetal2015}. 
By construction, $R^0_{\rm g}(\langle E\rangle)$ is appropriate
if it is known or presumed that the disordered conformations 
populate a narrow range of $R_{\rm g}$'s or distribute symmetrically
around an average $R_{\rm g}$~\cite{Songetal2015}, whereas 
$\sqrt{\langle R^2_{\rm g}\rangle(E)}$ is suitable when such
knowledge or assumption is absent. Therefore,
it is our contention that, given a single $\langle E\rangle_{\rm exp}$ 
{\it in the absence of additional experimental data}, the quantity
$\sqrt{\langle R^2_{\rm g}\rangle(E)}$ should serve well as the physically
valid Bayesian inference. However, if the $R_{\rm g}$'s are
known experimentally to be confined to a narrow range, 
which may be the case for certain IDPs, 
$R^0_{\rm g}(\langle E\rangle)$ would be the valid inference when
no further information besides $\langle E\rangle_{\rm exp}$ 
and the confinement is available.
The data provided in Fig.~6 and the Supporting
Information of ref.~\cite{Songetal2015} as well as those
in the present Figs.~3 and 6 are useful for this purpose.
\\

\begin{figure}[t]
\begin{center}
{\includegraphics[width=100mm,angle=0]{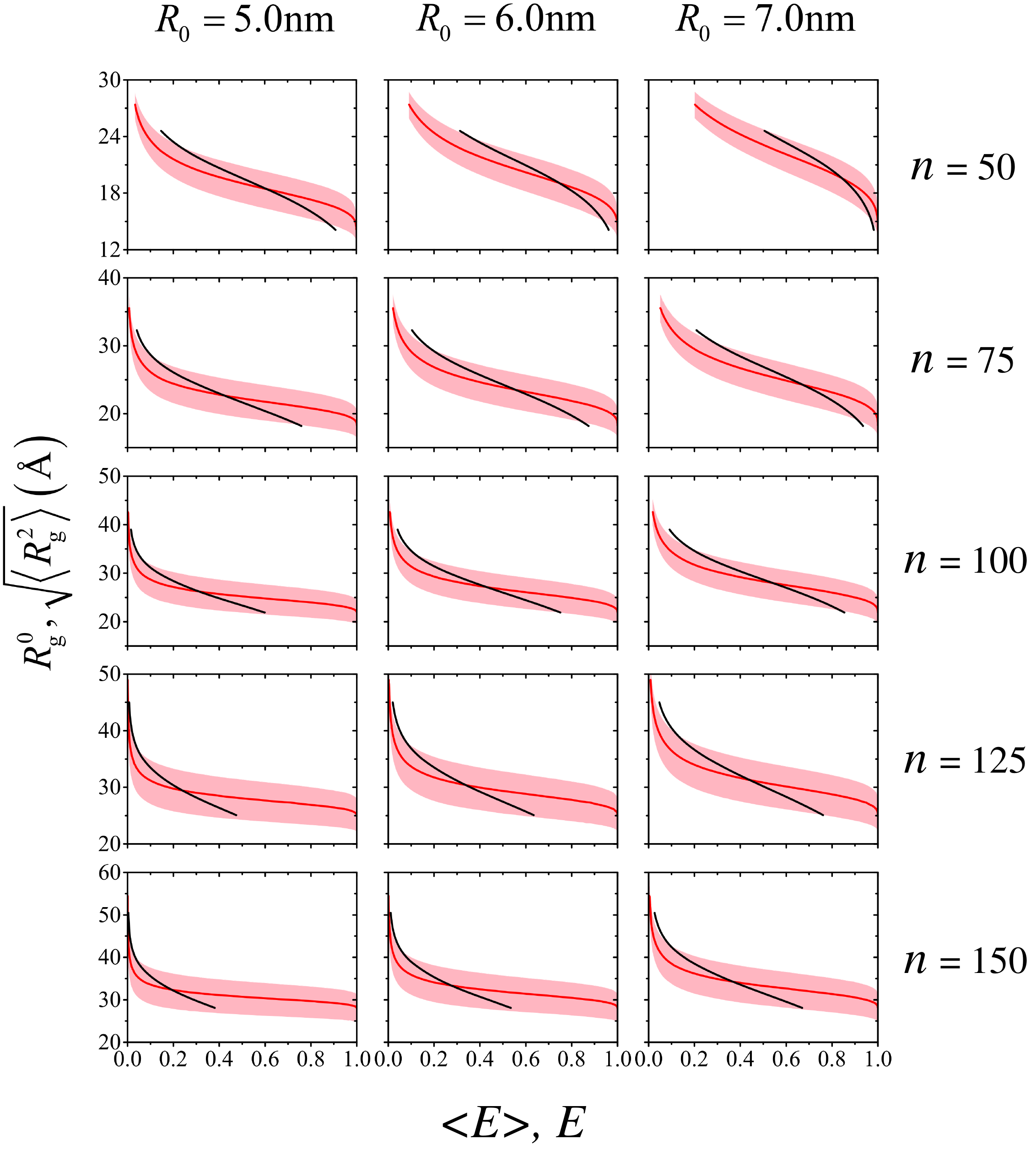}}
\vspace{0.0cm}
\end{center}
\caption{Most probable and root-mean-square radius of gyration.
Generalization of the $R^0_{\rm g}(\langle E\rangle)$ (solid black curves),
and $\sqrt{\langle R^2_{\rm g}\rangle(E)}$ (solid red curves)
for $R_0=55$ \AA~ and $n=75$ in Fig.~3
to other F\"orster radii $R_0$ and chain lengths $n$.
The shaded areas are bound by
$\sqrt{\langle R^2_{\rm g}\rangle(E) \pm \sigma(R^2_{\rm g})(E)}$,
which were represented by red dashed curves in Fig.~3.
As discussed in the text, the $\sqrt{\langle R^2_{\rm g}\rangle(E)}$
curves computed here for sharply defined $E$ values are expected to
apply also to $\sqrt{\langle R^2_{\rm g}\rangle(\langle E\rangle)}$ for
essentially symmetric distributions of $E$ where $\langle E\rangle$ denotes
the mean value of $E$ in such distributions.
As pointed out above for Fig.~3,
the black $R^0_{\rm g}(\langle E\rangle)$ curves shown here
do not cover $\langle E\rangle$ values close to zero or unity
because of the relatively large $R_{\rm g}$
bin sizes used previously~\cite{Songetal2015}.}
\end{figure}

{\bf Physically valid interpretation of smFRET data requires
explicit-chain modeling.} 
Conventional approaches to smFRET inference neglects possible
sequence-dependent conformational heterogeneity of unfolded
ensembles. They always enforce a full
conformational ensemble that expands or contracts 
homogeneously~\cite{haran2006,eaton2007}. Lacking an explicit-chain
representation, this elementary unphysicality of conventional smFRET
inference was often overlooked. Consequently, when
$\langle E\rangle_{\rm exp}$ is small, these procedures force the entire
ensemble to expand, leading to unrealistically high inferred 
$\langle R_{\rm g}\rangle$ values~\cite{Songetal2015}.
Although conformations with large $R_{\rm EE}$ (and hence small $E$ or 
$\langle E\rangle$) and large $R_{\rm g}$ are part of our
subensemble analysis (e.g. Fig.~2f), these rare conformations in
our simulations did not arise from physically unrealistic long Kuhn lengths 
or unrealistic intrachain repulsion as in conventional 
approaches~\cite{Songetal2015}. This is the fundamental 
reason why conventionally inferred $\langle R_{\rm g}\rangle$ values 
differ from those simulated using physical, explicit-chain
models~\cite{claudiu2016,Songetal2015,dt2009,Reddy2016,zhirong2016b}, and
that such simulations, for Sic1~\cite{Songetal2015} and 
Protein L~\cite{Reddy2016} for example, 
produced smaller variations in $\langle R_{\rm g}\rangle$ 
consistent with the limits prescribed by our subensemble SAW
analysis~\cite{Songetal2015} (Fig.~S1).

In this perspective, recent computational investigations using explicit-chain 
simulations to
rationalize smFRET data represent significant advances. These efforts include
a study on Protein L using a denaturant-dependent construct based on 
a native-centric G\=o-like sidechain potential~\cite{Reddy2016}
and an all-atom, explicit-water molecular dynamics study on 
ACTR and an R17 variant \cite{best2016,schuler2016}. In these studies,
the conformational heterogeneity of unfolded/disordered ensembles
encoded by amino acid sequences is taken into account either by a 
structure-specific G\=o-like potential~\cite{Reddy2016} or a transferrable 
atomic force field~\cite{best2016,schuler2016}. However, it should
be emphasized that commonly used force fields may not capture the high 
degrees of folding cooperativity observed for real 
proteins~\cite{chanetal2011}. In particular, in comparison with experiment,
the disordered conformational ensembles predicted by several atomic force 
fields are too 
compact~\cite{tobin_forcefields,DavidShaw2,sarah15,zhuqing2017}. 
Efforts to address
this shortcoming is underway~\cite{sarah17,best2017,shea2017}. For the case of 
Protein L, an earlier study~\cite{TaoPCCP} using a denaturant-dependent 
coarse-grained sidechain model similar to the one used in the recent study by 
Maity and Reddy~\cite{Reddy2016} suggests that, even with an essentially
native-centric potential, the model is insufficiently
cooperative vis-\`a-vis experiment. Specifically, the predicted chevron 
plot for Protein L has a folding-arm rollover~\cite{TaoPCCP}, which is
absent in experiment~\cite{Plaxco1999}. This behavior is related
to denaturant-dependent shifts in the positions of transition 
and unfolded states in the model~\cite{TaoPCCP}, which would likely lead 
to a reduction in $\langle R_{\rm g}\rangle$ with decreasing
[GuHCl]. We view these known limitations of current potentials
for protein folding simulation as part of the very puzzle underscored by 
the smFRET-SAXS discrepancy.  The crux of the matter is, if the degrees 
of folding cooperativity for some---albeit not all---proteins, such as 
Protein L, are indeed as high as envisioned by SAXS 
measurements~\cite{Plaxco1999}, why can't common force fields capture the
phenomenon~\cite{TaoPCCP}?

In lieu of attempting to provide an accurate model of sequence-specific 
interactions, our subensemble SAW approach to smFRET inference does not 
presume any particular model of sequence-dependent conformational 
heterogeneity. By itself, our approach merely establishes a perimeter 
for physically realizable conformational variation~\cite{Songetal2015}. 
The rationale is to let
experiment take precedence in uncovering the actual conformational
heterogeneity. In other words, $P(R^2_{\rm g}|E)$ is a baseline distribution
upon which any re-weighting of conformational population by sequence-specific 
effects is to be considered without prejudgement. Under this conceptual 
framework, we make no generalization as to whether conformational dimensions 
of disordered proteins would or would not increase with increasing denaturant 
concentration. Such a verdict has to be made on a case-by-case basis 
depending on the nature of available experimental information in addition 
to the limited structural constraint provided by smFRET. For example, 
our previous study indicates that the dimensions of IDP Sic1 
increases when [GuHCl] is increased from 1 M to 5 M~\cite{Songetal2015}.
A more recent in-depth study using smFRET, SAXS as well as other experimental 
probes and computation has demonstrated convincingly that conformational 
dimensions of the IDP ACTR and a destabilized mutant of globular protein R17 
increase upon increasing [GuHCl] or [urea]~\cite{best2016,schuler2016}.
It is of relevance, however, that unlike Protein L~\cite{Plaxco1999}, 
R17 is not a two-state folder as its chevron plot has a nonlinear 
unfolding arm~\cite{internalfriction2012}.
\\

{\bf A hypothetical scenario for the case of Protein L.} 
To make conceptual progress toward understanding
the Protein L unfolded state, we first put aside 
potential experimental artifacts that might be caused, for example, by
the sensitivity of $R_{\rm g}$ to the fitting range of
the Guinier analysis and the difficulty in obtaining low-denaturant SAXS
data~\cite{schuler2016}. For the following consideration, we assume
that the SAXS finding of an essentially denaturant-independent 
$\langle R_{\rm g}\rangle\approx 25$~\AA~(ref.~\cite{Plaxco1999}) and the 
smFRET data of a decreasing $\langle E\rangle_{\rm exp}$ with increasing 
denaturant~\cite{haran2006,eaton2007} are both valid. We then
seek to rationalize the experimental data by constructing
denaturant-dependent heterogeneous conformational ensembles consistent 
with both sets of data. In so doing, we are merely following an 
investigative logic commonly 
practised in the construction of putative unfolded and 
IDP ensembles~\cite{schuler2016,julie2001,marsh2012,antonov2016}. 
As explained below, a solution to the smFRET-SAXS puzzle is
possible if, with decreasing denaturant, sequence-specific effects 
become increasing biased to re-distribute conformational population 
to high $R^2_{\rm g}$ values such that
a nearly constant $\sqrt{\langle R^2_{\rm g}\rangle}\approx 25$~\AA~is 
maintained despite the shift of the baseline Bayesian distribution 
$P(R^2_{\rm g}|\langle E\rangle)$ to lower $R^2_{\rm g}$ values 
because of increasing $\langle E\rangle_{\rm exp}$ 
with decreasing denaturant (Fig.~4).

How biased does such a denaturant-dependent conformational heterogeneity
need to be? Using the example in Fig.~4 for
unfolded Protein L at [GuHCl] = 1 M and 7 M, an estimate of
the necessary denaturant-dependent bias needed to resolve the 
smFRET-SAXS puzzle can be made.
Consider the Bayesian distributions $P(R^2_{\rm g}|E)$ (Fig.~4c) and 
$P(R^2_{\rm g}|\langle E\rangle)$ (Fig.~4d). These are baseline
distributions that do not account for any sequence-specific effect.
They show that $\approx 10\%$ and $\approx 25\%$, 
respectively, of the $E, \langle E\rangle_{\rm exp}\approx 0.74$ and 
$E, \langle E\rangle_{\rm exp}\approx 0.45$ populations have
$R_{\rm g} \ge 25$~\AA~($R^2_{\rm g} \ge 625$~\AA$^2$).
This means that different subsets of these two conformational distributions
can have the SAXS-observed $\sqrt{\langle R^2_{\rm g}\rangle}\approx 25$~\AA.
Indeed, possible sequence-specific re-weighted distributions for Protein L
that are consistent with both smFRET and SAXS may
take the forms of the shaded symmetric regions in Fig.~7
(grey, and pink plus grey areas). 
These distributions are consistent with both smFRET and SAXS 
because they both have $\sqrt{\langle R^2_{\rm g}\rangle}\approx 25$~\AA~(thus
consistent with SAXS) yet $\langle E\rangle \approx 0.74$ 
($\langle E\rangle_{\rm exp}$ at [GuHCl] = 1 M) for the grey distribution and
$\langle E\rangle \approx 0.45$ ($\langle E\rangle_{\rm exp}$ at [GuHCl] = 7 M) 
for the pink plus grey distribution. 

That this holds true is easy to see if the distributions in
question are for two sharply defined $E$'s. In that case, we use
the two $P(R^2_{\rm g}|E)$'s in Fig.~4c to define
two restricted (unnormalized) distributions $P_{\rm r}(R^2_{\rm g}|E)$ 
such that $P_{\rm r}(R^2_{\rm g}|E)=P(R^2_{\rm g}|E)$ for 
$R^2_{\rm g}\ge 625$~\AA$^2$ and
$P_{\rm r}(R^2_{\rm g}|E)=
\min[P(R^2_{\rm g}|E),P(\{2\times 625{\rm \AA}^2-R^2_{\rm g}\}|E)]$ 
for $R^2_{\rm g}< 625$~\AA$^2$.
Because of the mirror symmetry of these distributions with respect to
$R^2_{\rm g}=625$~\AA, the values of their
$\sqrt{\langle R^2_{\rm g}\rangle}=$
$[\int dR^2_{\rm g}\; R^2_{\rm g}\; P_{\rm r}(R^2_{\rm g}|E)]^{1/2}$
are both $\approx 25$~\AA~even though $E=0.447$ for all conformations in 
the $P_{\rm r}(R^2_{\rm g}|E=0.45)$ distribution and $E=0.745$ for 
all conformations in the $P_{\rm r}(R^2_{\rm g}|E=0.75)$ distribution.
This result is generalizable
to the two broad $P(E)$ distributions in Fig.~4b.
Consider $\int dE\; P(E) P_{\rm r}(R^2_{\rm g}|E)$.
By definition this integral 
gives exactly the $R^2_{\rm g} \ge 625$~\AA$^2$ parts (in darker shades) 
of the grey, and pink plus grey areas in Fig.~7 because 
$P_{\rm r}(R^2_{\rm g}|E)=P(R^2_{\rm g}|E)$ for
$R^2_{\rm g}\ge 625$~\AA$^2$ and
$P(R^2_{\rm g}|\langle E\rangle) = \int dE\; P(E) P(R^2_{\rm g}|E)$.
The integral yields close approximations to the $R^2_{\rm g} < 625$~\AA$^2$ 
lighter shaded areas in Fig.~7 because 
$\sqrt{\langle R^2_{\rm g}\rangle(E)}$ varies mildly 
in the range $0.2 \le E \le 0.95$ (Fig.3b) that 
covers most of the $P(E)$ distributions (Fig.~4b). 
This procedure ensures that the conformational populations 
represented by the grey plus pink and grey areas 
in Fig.~7 preserve
their respective $\langle E\rangle = \int dE\; EP(E)$ values because 
$\int dE\; P(E) P_{\rm r}(R^2_{\rm g}|E)$ preserves
the average $E$ at every $R^2_{\rm g}$.
Therefore, the shaded distributions in Fig.~7 
represent conformations with
different $\langle E\rangle \approx 0.45$ and 
$\langle E\rangle \approx 0.74$ but possess the same 
$\sqrt{\langle R^2_{\rm g}\rangle}\approx 25$~\AA.
This hypothetical scenario indicates that consistency between
SAXS and smFRET is possible if sequence-induced heterogeneity
entails a mild restriction to $\sim 2\times 25\% = 50\%$ of 
the conformational possibilities allowed by the
$\langle E\rangle_{\rm exp}$ at [GuHCl] = 7 M but imposes a more severe 
restriction to $\sim 2\times 10\% = 20\%$ of the conformational possibilities 
allowed by the $\langle E\rangle_{\rm exp}$ 
at [GuHCl] = 1 M (Fig.~7). It should be emphasized, however, that
this is only one among many
possible scenarios of denaturant-dependent conformational re-weighting
that can satisfy both smFRET and SAXS data. 
Further information about the re-weighting may be offered by
additional experimental data such as pair distributions from SAXS, but
that is beyond the scope of this work.

\begin{figure}[t]
\begin{center}
{\includegraphics[width=130mm,angle=0]{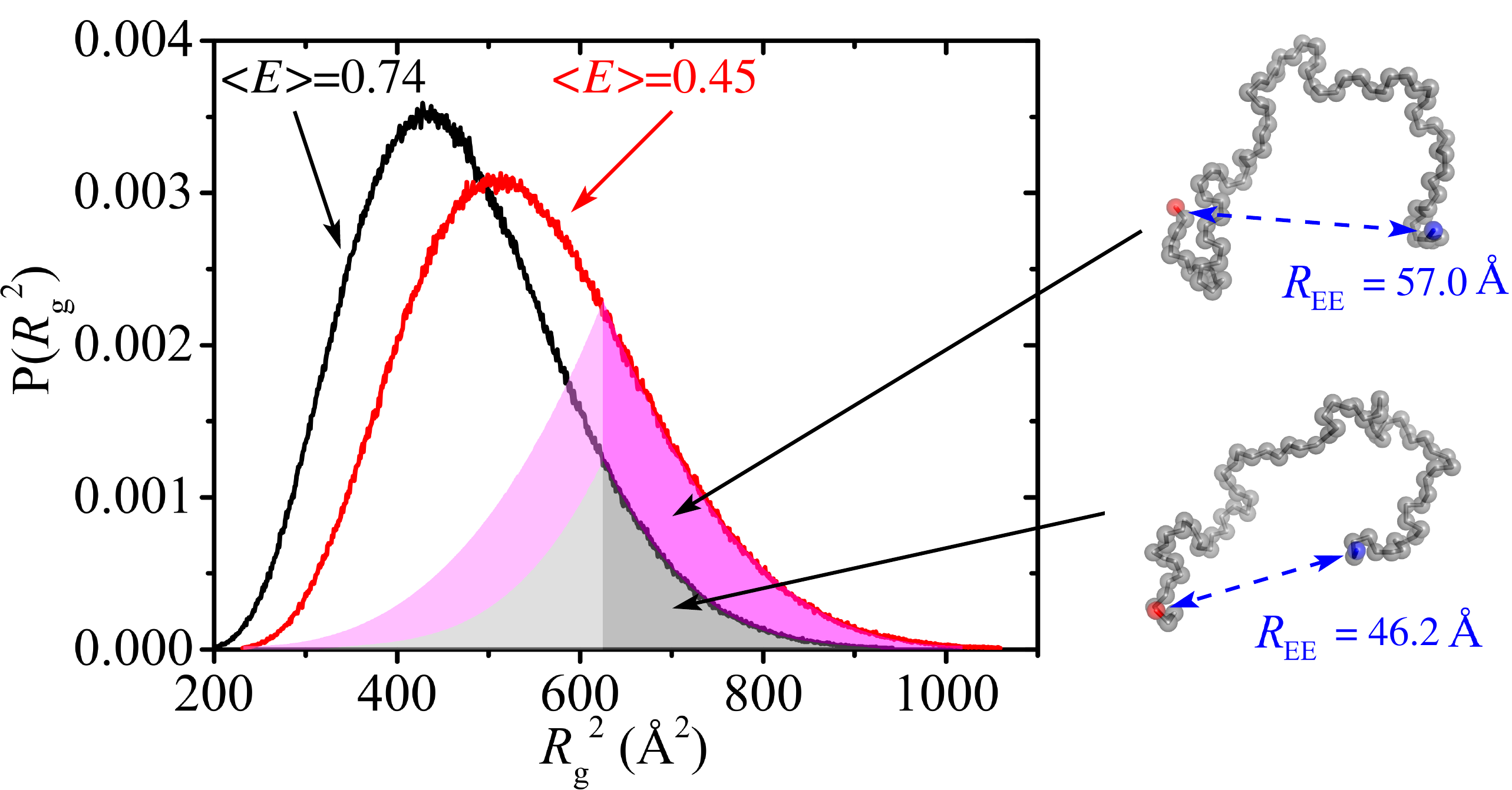}}
\vspace{0.0cm}
\end{center}
\caption{A hypothetical resolution of the Protein L
smFRET-SAXS puzzle. The two distributions depicted by the black and red
curves are from Fig.~4d, for $\langle E\rangle=0.74$ and
$\langle E\rangle=0.45$, respectively. For $R^2_{\rm g}\ge 625$ \AA,
area shaded in pink is under the $\langle E\rangle=0.45$ (red)
distribution but above the $\langle E\rangle=0.74$ (black) distribution,
whereas area shaded in grey is under the $\langle E\rangle=0.74$ (black)
distribution. The $R^2_{\rm g}< 625$~\AA~areas that are in lighter
shades are mirror reflections of the corresponding
$R^2_{\rm g}\ge 625$ \AA~areas with respect to $R^2_{\rm g}=625$~\AA.
The sumtotal of the pink plus grey area 
($\sim 50\%$ of $P(R^2_{\rm g}|\langle E\rangle=0.45)$)
represents a hypothetical ensemble with
$\langle E\rangle\approx 0.45$ and $\sqrt{R^2_{\rm g}}\approx 25$~\AA,
whereas the grey area ($\sim 20\%$ of $P(R^2_{\rm g}|\langle E\rangle=0.74)$)
represent a hypothetical ensemble with $\langle E\rangle\approx 0.74$ but 
nonetheless the same $\sqrt{R^2_{\rm g}}\approx 25$~\AA.
Shown on the right are example
conformations in these restricted ensembles, as marked by the arrows.
Both conformations
have $R^2_{\rm g} = 700$~\AA$^2$ ($R_{\rm g}=26.5$~\AA), but their
different $R_{\rm EE}$ values entail different $E$ values of $\approx 0.45$
(top) and $\approx 0.74$ (bottom). See text and Fig.~4 for further details.}
\end{figure}

The denaturant-dependent biases represented by the above estimates
are intuitively plausible because the required biases of
$50\% \rightarrow 20\%$ for [GuHCl] = 7 M $\rightarrow$ 1M are not excessive.
These fractional restrictions are only 
rough estimates, but they serve to illustrate a key concept. 
It is conceivable that the required restrictions can be less.
For instance, when the atomic size and shapes of amino acid sidechains
are taken into account, the actual intraprotein excluded volume effect can
be stronger than that embodied by the $R_{\rm hc}=4$~\AA~repulsion distance
in the C$_\alpha$ model. If $R_{\rm hc}=5$~\AA~is used 
instead~\cite{Songetal2015}, the $R_{\rm g}$ distribution would shift 
upward by $\approx 1$--$3$~\AA~(Fig.~S3). In that case, the 
fractions of $P(R^2_{\rm g}|\langle E\rangle)$ with 
$R_{\rm g} \ge 25$~\AA~would increase, enabling significantly less 
severe denaturant-dependent biases of $81\% \rightarrow 43\%$
(for [GuHCl] = 7 M $\rightarrow$ 1M)
to resolve the smFRET-SAXS discrepancy (Fig.~S4).
\\

{\bf Concluding remarks.}
We deem this scenario for Protein L viable pending further experiment 
because natural proteins 
are heteropolymers, not homopolymers. Their amino acid sequences encode
for heterogeneous intrachain interactions, especially under strongly folding 
(low or zero denaturant) conditions, which logically can only lead to 
heterogeneous conformational ensembles even when the chains are disordered.
Unfolded conformations are not Gaussian chains~\cite{topomer2005}.
The question is not whether heterogeneity exists but the degree 
of heterogeneity and its impact. 
Such heterogeneity is observable by NMR~\cite{baldwin1995}, in some cases 
even in high urea concentrations~\cite{shortle2001,DanRohit2013},
not only for proteins such as BBL that do not fold 
cooperatively~\cite{munoz2006}, but also for two-state folders
(as defined by equality of van't Hoff and calorimetric enthalpies of
unfolding, and chevron plots with linear folding
and unfolding arms~\cite{chanetal2011,chanetal2004}) such as
cytochrome c~\cite{bai1995}. The biophysics of protein folding processes
that are macroscopically cooperative yet microscopically heterogeneous is
readily understood theoretically~\cite{shimizu2002,kaya2005,knott06}.
From a mathematical standpoint, it is definitely possible, as we envisioned
above, for heterogeneous conformational ensembles that are distinct from 
random coils or SAWs to have overall random-coil or SAW 
dimensions nonetheless~\cite{Songetal2015}, 
as has been demonstrated by a recent study of 
the IDP Ash1~\cite{martin2016} and by hypothetical explicit-chain 
ensembles constructed to embody such properties~\cite{rose2000,rose2004}.
The scenario we suggested for resolving the smFRET-SAXS discrepancy
for Protein L posits an increased population of transient loop-like 
disordered conformations with the two chain termini close to each other 
under native conditions. Is this feasible? Of relevance to this
question is the experimental finding that conformations 
with enhanced populations of nonlocal contacts are
involved in the folding kinetics of adenylate 
kinase~\cite{haas2009,haas2014,haas2016}. Conformations with similar
properties have also been suggested by theory to be favored along folding 
transition paths~\cite{zhang12}. 
Recently, a disordered conformational state with 
such properties was identified for the protein drkN SH3 as well, 
though in this case it is induced by high rather than by low 
denaturant~\cite{claudiu2016}. All in all, it is clear from the above
considerations that denaturant-dependent heterogeneity 
in disordered protein conformational ensembles is expected in general. 
To what degree and in what manner it may account for the smFRET-SAXS 
discrepancy will have to be ascertained by further experiment.
\\

Recently, Fuertes et al.~[94] make an observation
similar to ours---among other results of theirs---that 
the smFRET-SAXS puzzle may be resolved by recognizing 
that a given $R_{\rm EE}$ can be consistent with a variety of $R_g$ values. 
For the record, it is noted that one of the authors of this 
work~\cite{lemke2017} kindly sent their manuscript (submitted but 
unpublished at the time) 
to us after we shared with him our paper on May 15, 2017 
before submitting the original version of the present paper to this journal 
and making it publicly available on arXiv.org (arXiv:1705.06010).

$\null$\\
{\bf\large Supporting Material}\\
Supporting Information comprises four supporting figures is available 
at the {\it Biophysical Journal} website.

$\null$\\
{\bf\large Author Contributions}\\
J.S. and H.S.C. designed the research.
J.S., G.-N.G. and H.S.C. performed the research.
J.S., G.-N.G., C.C.G. and H.S.C. analyzed the data.
T.S. contributed computational tools.
J.S. and H.S.C. wrote the paper.

%%%%%%%%%%%%%%%%%%%%%%%%%%%%%%%%%%

$\null$\\
{\large\bf Acknowledgments}\\ 
H.S.C. thanks Osman Bilsel, Kingshuk Ghosh, Elisha Haas, Rohit Pappu,
and Tobin Sosnick for helpful 
discussions during Protein Folding Consortium workshops sponsored by 
the National Science Foundation (US), and Eitan Lerner for comments on 
an earlier version of this paper. J.S. gratefully acknowledges support from
the National Natural Science Foundation of China (Grant No. 21674055) and
the Open Research Fund of State Key Laboratory of Polymer Physics 
and Chemistry, Changchun Institute of Applied Chemistry, 
Chinese Academy of Sciences (Grant No. 201613). G.-N.G. was supported 
by an Ontario Graduate Scholarship. Support for this work was also 
provided by Natural Science and Engineering Research Council of Canada 
Discovery Grant RGPIN 342295-12 to C.C.G., Canadian Institutes of Health 
Research Operating Grant No. MOP-84281 to H.S.C., and generous 
allotments of computational resources from SciNet of Compute/Calcul Canada.

%\input FRETcaps1
%\vfill\eject

%%%%%%%%%%%%%%%%%%
\vfill\eject

\centerline{\Huge Supporting Information}

$\null$

\centerline{\large\it for}

$\null$

\centerline{\large{\it Biophysical Journal} article}
\vskip 0.3in

\begin{center}
{\Large\bf Conformational Heterogeneity and FRET Data}\\

\vskip 0.1in

{\Large\bf Interpretation for Dimensions of Unfolded Proteins}\\

\vskip .5in
{\bf Jianhui S{\footnotesize{\bf{ONG}}}},$^{1,2}$
{\bf Gregory-Neal G{\footnotesize{\bf{OMES}}}},$^{3}$\\
{\bf Tongfei S{\footnotesize{\bf{HI}}}},$^{4}$
{\bf Claudiu C. G{\footnotesize{\bf{RADINARU}}}},$^{3}$
 and
{\bf Hue Sun C{\footnotesize{\bf{HAN}}}}$^{2,*}$

$\null$\\
$^1$ School of Polymer Science and Engineering,
Qingdao University of\\ Science and Technology,
53 Zhengzhou Road, Qingdao 266042, China;\\
$^2$ Departments of Biochemistry and Molecular Genetics,\\
University of Toronto, Toronto, Ontario M5S 1A8, Canada;\\
$^3$ Department of Chemical and Physical Sciences,\\
University of Toronto Mississauga, Mississauga, Ontario L5L 1C6 Canada; and\\
Department of Physics, University of Toronto, Toronto, Ontario M5S 1A7,
Canada;\\
$^4$ State Key Laboratory of Polymer Physics and Chemistry,\\
Changchun Institute of Applied Chemistry, Chinese Academy of Sciences,\\
Changchun 130022, China

\vskip .5in

\end{center}
\noindent
$^*$
Corresponding author.\\
{\phantom{$^*$\ }}Hue Sun Chan.
E-mail: {\tt chan@arrhenius.med.toronto.edu}\\

\vfill\eject

\begin{center}
{\LARGE\bf Supporting Figures}
$\null$\\
$\null$\\
{\includegraphics[height=90mm,angle=0]{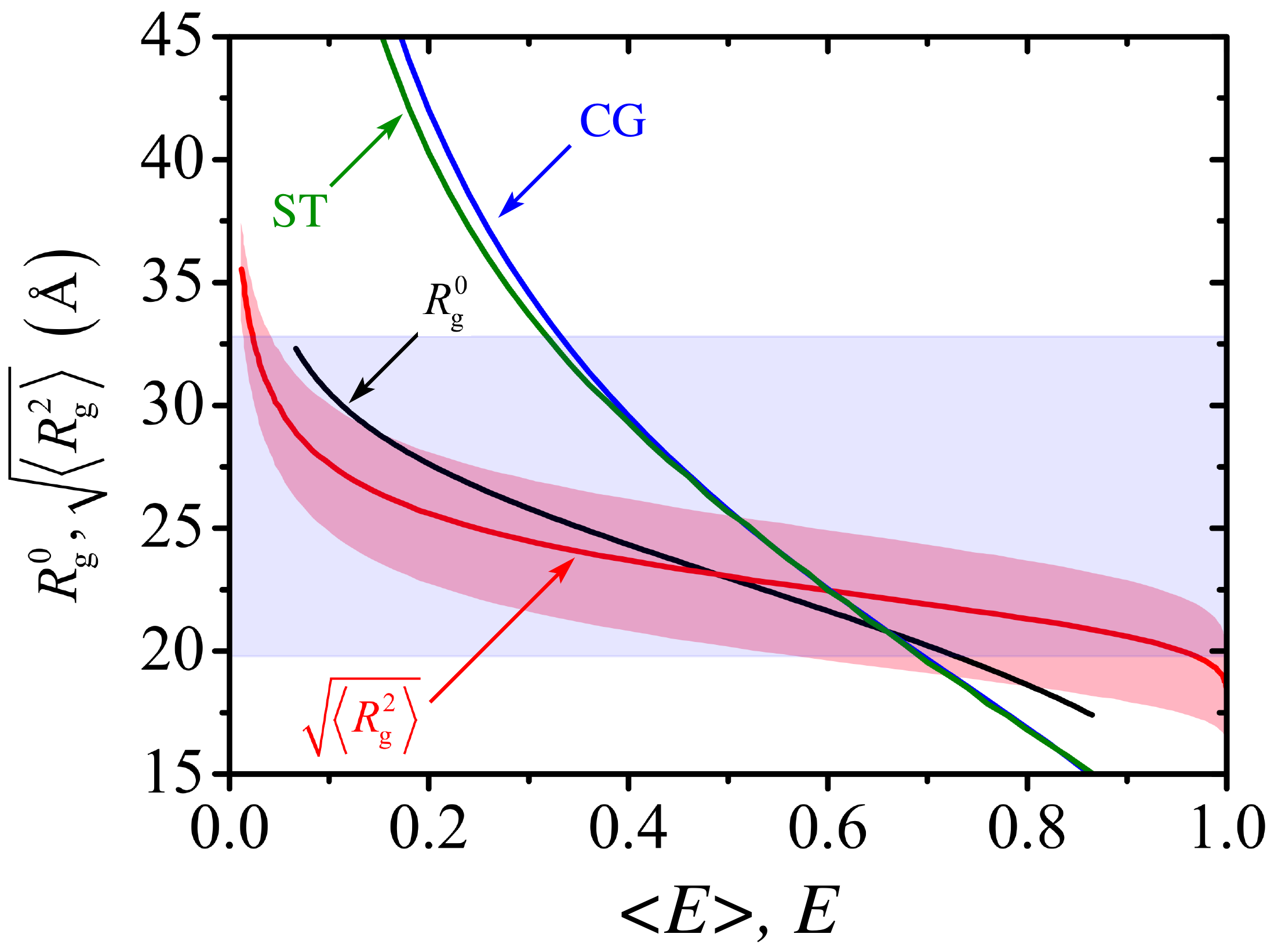}}
\end{center}
\vspace{0.0cm}
%\caption{}
%\label{}
$\null$\\
{\large\bf Figure S1.} Comparing subensemble-based and conventional
smFRET inferences of conformational dimensions.
The most probable $R^0_g(\langle E\rangle)$ (black curve) and the 
root-mean-square $\sqrt{\langle R^2_g\rangle}(E)$
(red curve) for $n=75$ and $R_0=55$ \AA~are the same as those
in Fig.~3 of the main text. The pink-shaded area here corresponds to 
the area bounded by the red dashed curves in Fig.~3 of the main text
for $\sqrt{\langle R^2_g\rangle \pm \sigma(R^2_g)}$. Included for
comparison are conventional smFRET inference using either the 
Gaussian chain (GC, blue curve) or the Sanchez theory (ST, green curve) 
methods as described previously [Song, J., G.-N. Gomes, C. C. 
Gradinaru, and H. S. Chan. 2015.  An adequate account of excluded volume 
is necessary to infer compactness and asphericity of disordered proteins 
by F\"orster resonance energy transfer. {\it J. Phys. Chem. B} 
119:15191--15202]. As is clear from Fig.~6 of this reference and
also in the present figure, conventional smFRET inference methods of 
CG and ST posit a much sharper variation in inferred radius of gyration 
as a function of average transfer efficiency $\langle E\rangle$. 
The light blue area ($19.79$~\AA~$\le R_g \le 32.80$~\AA) marks the
range of expected radii of gyration for fully unfolded protein
ensembles with chain length $n=75$ as provided by Kohn et al.
[Kohn, J. E., I. S. Millett, J. Jacob, B. Zagrovic, T. M. Dillon,
N. Cingel, R. S. Dothager, S. Seifert, P. Thiyagarajan, T. R. Sosnick,
M. Z. Hasan, V. S. Pande, I. Ruczinski, S. Doniach, and K. W. Plaxco. 2004.
Random-coil behavior and the dimensions of chemically unfolded proteins.
{\it Proc. Natl. Acad. Sci. USA} 101:12491--12496].

\vfill\eject

\begin{center}
{\includegraphics[height=59mm,angle=0]{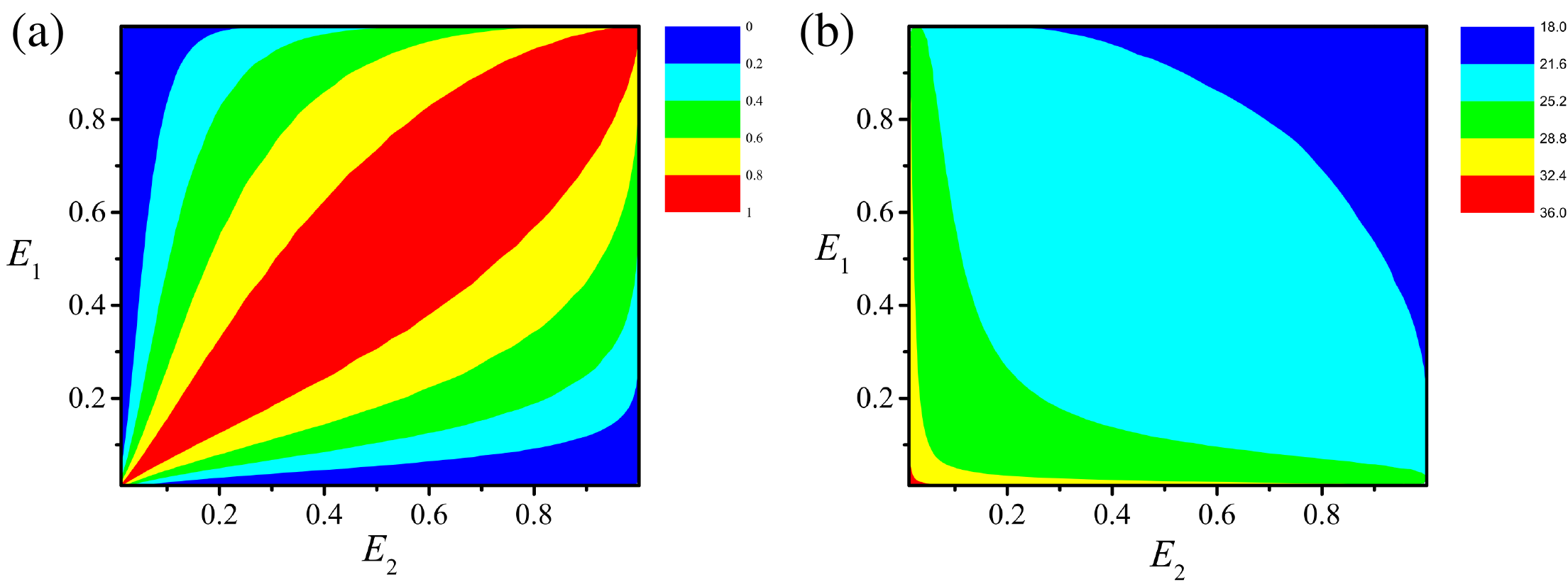}}
\end{center}
\vspace{0.0cm}
%\caption{}
%\label{}
$\null$\\
{\large\bf Figure S2.}\\
Overlapping $R^2_{\rm g}$ distributions for pairs of FRET efficiencies.
Results shown are for $n=75$ and $R_0=55$ \AA.
(a) Same data as Fig.~5 of the main text plotted in a different
style. The color code here indicates range of values for the
overlapping coefficient ${\rm OVL}[P(R^2_{\rm g}|E_1),P(R^2_{\rm g}|E_2)]$.
The fractional areas in red, yellow, green, cyan, and blue are,
respectively, 0.311, 0.267, 0.193, 0.128, and 0.101.
(b) Root-mean-square radius of gyration averaged over the overlapping
region of $P(R^2_{\rm g}|E_1)$ and $P(R^2_{\rm g}|E_2)$. The value
represented by the color code is given by
$\sqrt{\int dR^2_{\rm g} \; R^2_{\rm g}
\{\min [P(R^2_{\rm g}|E_1),P(R^2_{\rm g}|E_2)]\}}$.
For instance, this quantity for the pair of distributions in Fig.~4c of
the main text with $E_1\approx 0.447$ and
$E_2\approx 0.745$ (OVL $=0.747$) is equal to
$\sqrt{503.6 {\rm \AA}^2}=22.4$ \AA. Note that this value is practically
identical to the value of $\sqrt{505.1 {\rm \AA}^2}=22.5$ \AA~for the
root-mean-square radius of gyration averaged over the overlap area in
Fig.~4d of the main text for two broad $E$ distributions with OVL = 0.754.

\vfill\eject

$\null$

\begin{center}
\vskip -1.5cm
{\includegraphics[height=130mm,angle=0]{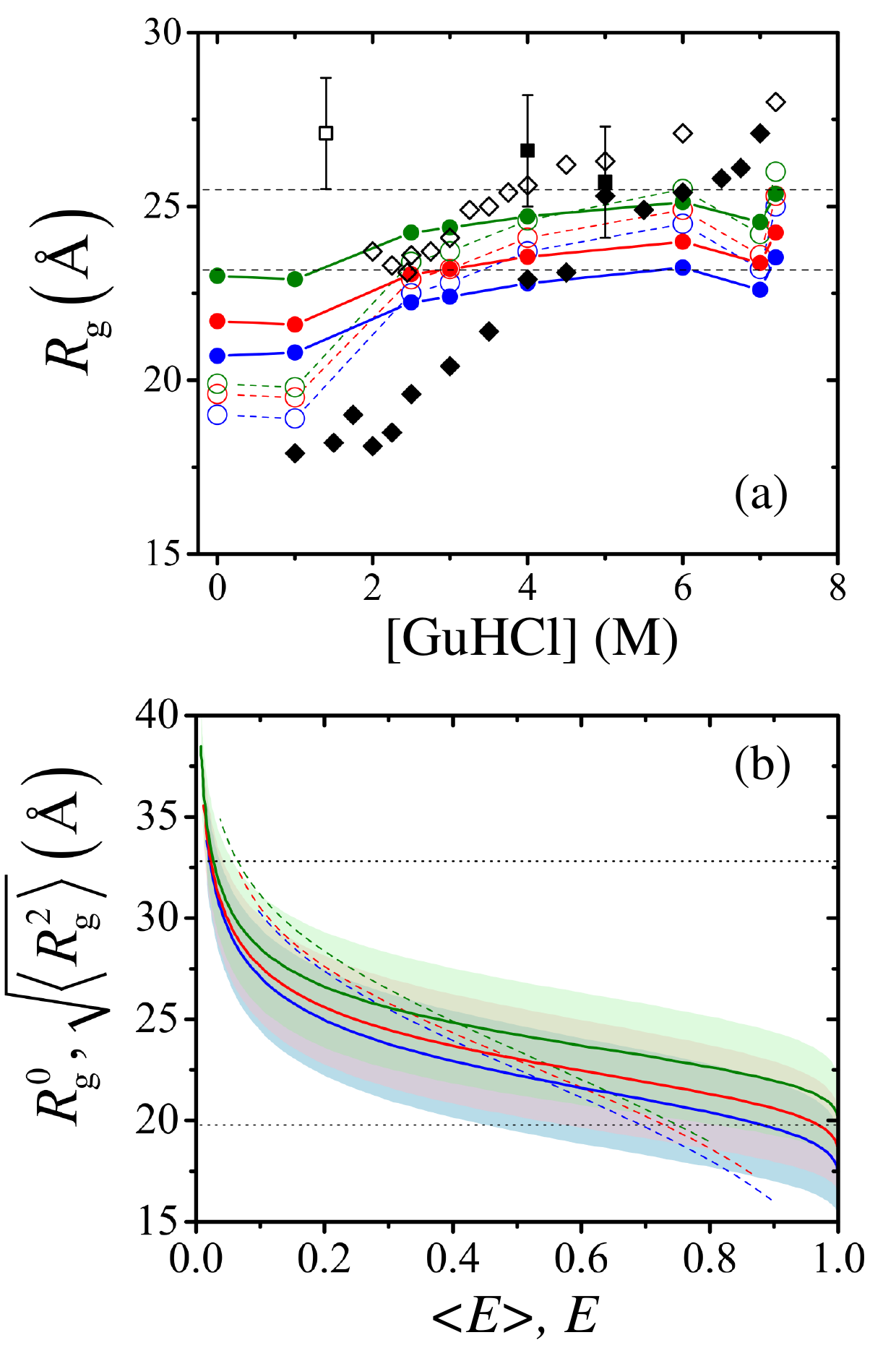}}
\end{center}
\vspace{0.0cm}
%\caption{}
%\label{}
$\null$\\
{\large\bf Figure S3.} Variation in subensemble-based smFRET inference
due to differences in assumed intraprotein excluded volume.
(a) is based on Fig.~1 of the main text. The black squares and
diamonds (SAXS data) as well as the open red circles ($R^0_g$) and 
filled red circles ($\sqrt{\langle R^2_g\rangle}$) for 
hard-core repulsion distance $R_{\rm hc}=4.0$~\AA~have the same meanings
as the corresponding symbols in Fig.~1 of the main text.
The other circular symbols here also represent $R^0_g$ and 
$\sqrt{\langle R^2_g\rangle}$ but are for 
$R_{\rm hc}=3.14$~\AA~(green) and $R_{\rm hc}=5.0$~\AA~(blue).
Error bars showing spreads in the $P(R^2_g|E)$ distributions 
are not shown. The dashed and solid lines connecting the circular symbols are
merely guides for the eye. The two horizontal dashed black lines 
indicate the expectation  by Kohn et al. (referenced in Fig.~S1)
for $R_g = 25.48$~\AA~when $n=75$ (length of Protein L plus dye linkers)
and $R_g = 23.17$~\AA~when $n=64$ (length of Protein L itself).
(b) $R^0_g(\langle E\rangle)$ (dashed curves) and 
$\sqrt{\langle R^2_g\rangle}(E)$ (solid curves) for 
$R_{\rm hc}=4.0$~\AA~(red, same as in Fig.~3b of the main text),
$R_{\rm hc}=3.14$~\AA~(green) and $R_{\rm hc}=5.0$~\AA~(blue); all for
$n=75$ and $R_0=55$~\AA. The areas bounded by the corresponding
$\sqrt{\langle R^2_g\rangle \pm \sigma(R^2_g)}$'s 
are shaded in the same colors with transluency indicating their overlaps.
The two horizontal dashed lines mark the
$19.79$ and $32.80$~\AA~boundaries in Fig.~S1 of
the expected $R_g$ range for fully unfolded $n=75$ ensembles.

\vfill\eject

\begin{center}
{\includegraphics[height=110mm,angle=0]{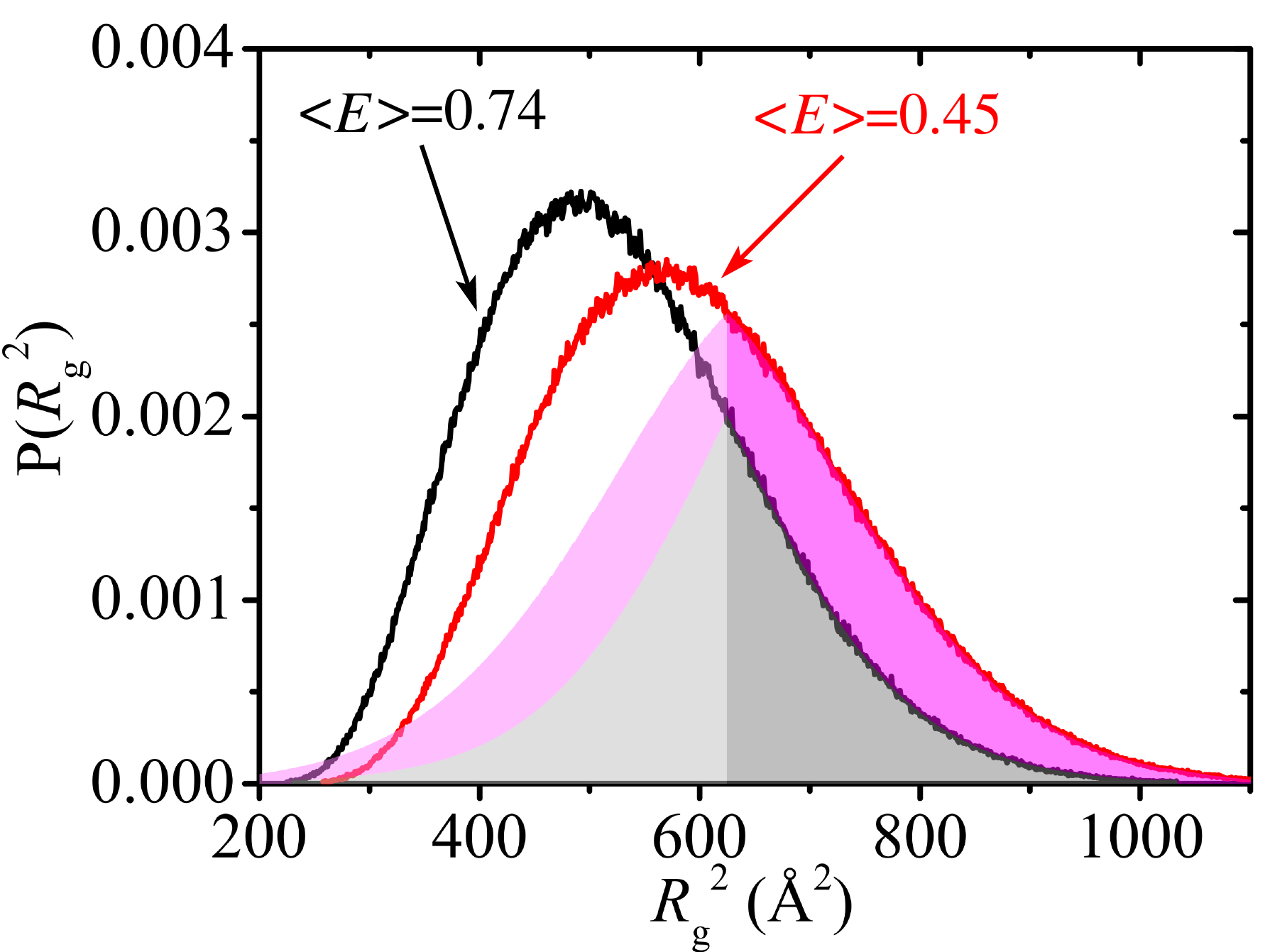}}
\end{center}
\vspace{0.0cm}
%\caption{}
%\label{}
$\null$\\
{\large\bf Figure S4.} A scenario in which less denaturant-dependent 
conformational bias would be needed to resolve the smFRET-SAXS puzzle of 
Protein L if enhanced intraprotein excluded volume effects are assumed. 
Simulation data conveyed by the present figure for $n=75$ and
$R_0=55$~\AA~are the same
as those in Fig.~7 of the main text except here $R_{\rm hc}=5.0$~\AA~instead
of the $R_{\rm hc}=4.0$~\AA~in that figure.
As in the main-text figure, the present black and red $P(R^2_g)$ 
distributions (OVL = $0.782$) are for the two $P(E)$ distributions 
of transfer efficiencies shown in Fig.~4b of the main text. 
Now the grey-shaded area makes up $43\%$ of the black $P(R_g^2)$ 
distribution, whereas the sum of the grey-shaded and pink-shaded
areas constitutes $81\%$ of the red $P(R^2_g)$ distribution.

\vfill\eject

\end{document}